\documentclass{article}

\usepackage[english]{babel}  
\usepackage[top=2cm, bottom=3cm, left=2cm, right=2cm]{geometry}
\usepackage{graphicx}
\usepackage{amsmath}
\usepackage{amssymb}
\usepackage{upgreek}
\usepackage{chemmacros} 
\usepackage{hyperref}
\hypersetup{
  colorlinks=true,
  linkcolor=blue,
  urlcolor=blue,
  filecolor=blue,
  citecolor=blue
}
\usepackage{cleveref}
\usepackage{float}
\usepackage{grffile}
\usepackage{txfonts}
\usepackage{booktabs}
\usepackage{dsfont}
\usepackage{subcaption}
\usepackage{array}
\usepackage{multirow}

\title{Numerical investigation of engine position effects on contrail formation and evolution in the near-field of a realistic aircraft configuration}
\author{}
\date{}

\begin{document}
\maketitle

\noindent {\large Rémy Annunziata$^{\star}$, Nicolas Bonne$^{\star}$, François Garnier $^\ddag$}\\

\noindent \textbf{Abstract:} The present study investigates the impact of engine position on contrail formation and near-field evolution in a realistic three-dimensional aircraft configuration. Detailed numerical simulations are conducted using a Reynolds-Averaged Navier-Stokes (RANS) approach coupled with mesh adaptation techniques. A Eulerian microphysical model is used to characterize contrail ice crystal properties and their evolution under varying dilution conditions. The setup is based on a Boeing 777-like geometry, including fuselage, wings, engines, and tailplane. Two microphysical activation scenarios are considered: one incorporating adsorption-based ice nucleation and the other assuming fully activated soot particles. The latter for two soot number emission indices. The dilution process and wake structure exhibit a strong dependence on engine placement, which significantly influences plume saturation. In highly diluted configurations, enhanced early-stage mixing reduces plume temperature and increases relative humidity, favoring the growth of larger ice crystals. Depending on the soot number concentration, vapor depletion effects may outweigh dilution-driven changes in water vapor availability. In adsorption-limited activation scenarios, increased dilution reduces the concentration of sulfur species, leading to a lower activation fraction and the formation of smaller ice crystals. Additionally, across the scenarios, the modified jet-vortex interaction alters particle distribution and their access to water vapor, further shaping their growth. These effects ultimately impact the contrail’s optical properties, particularly its optical thickness.\\


\noindent \rule{\linewidth}{.5pt}

\section{Introduction}
\label{sec:section1}

\subsection{Environmental impact and contrail formation}

Condensation trails (contrails), the linear white clouds formed in the wake of aircraft, constitute a significant non-\ch{CO2} contribution of aviation to environmental degradation \cite{LEE2009}. Under favorable atmospheric conditions, particularly in regions of high humidity, contrails can transition into cirrus-like clouds that may persist in the atmosphere for several hours. During this period, contrails affect the Earth's radiation balance by reflecting incoming solar radiation and trapping outgoing terrestrial radiation, thereby contributing to global radiative forcing \cite{BURK2011} and global warming. Their estimated contribution to radiative forcing is twice that of \ch{CO2} \cite{LEE2021}. However, this estimate is associated with significant uncertainties, underscoring the need for a more comprehensive understanding of the mechanisms governing their formation and evolution. Such knowledge is essential for accurately evaluating their environmental impact and developing strategies, whether by preventing their formation or minimizing their atmospheric persistence. In this perspective, investigating the process of contrail formation becomes crucial, as it primarily dictates their lifetime \cite{UNTER2014} and optical properties, both of which are closely related to their radiative impact \cite{KAR2018}.

\subsection{Numerical approaches to contrail modeling}

Contrail formation is commonly investigated through modeling and numerical simulations. These models range in complexity, addressing both microphysical and dynamical aspects of contrail development. Regarding microphysics, the most detailed approaches are 0D box models that simulate particle trajectories \cite{KAR1998,WONG10,BIER22}. While these models accurately characterize microphysical processes and chemical kinetics, their treatment of dynamics relies on a semi-empirical law for plume dilution \cite{SCHU98} or on path-lines derived from dynamic simulation \cite{BIER22}. As a result, they cannot fully represent the complex dynamics and intense turbulence present in the aircraft wake, which is crucial for particle dispersion and water uptake. Alternative approaches, such as Large Eddy Simulation (LES), have been applied to study contrail formation \cite{PAOLI2003,PAOLI2005,PAO2013}. LES methods initialize a jet and a vortex, allowing for the simulation of turbulent mixing and the intricate interactions between the jet and the wingtip vortex in the aircraft wake. These studies highlight the role of jet-vortex interactions in contrail formation. In addition, Lewellen \cite{Lewe20} demonstrates, through comparisons between LES and box models, that under conditions of elevated aerosol concentrations and inhomogeneous turbulent mixing, box models exhibit limitations in accurately representing contrail ice crystal number concentrations.

With recent advances in computational capabilities, methods have been developed that incorporate aircraft geometry, enabling the simulation of flow around the aircraft. These approaches can account for both mixing phenomena and the intense turbulence generated in the wake. These 3D approaches rely on Reynolds-Averaged Navier-Stokes (RANS) modeling. Guignery et al. \cite{GUIGNERY12} have demonstrated the ability of a RANS approach to simulate contrail formation using a NACA0012 wing. This work was later extended by Khou et al. \cite{Khou15,KHOU2016}, who successfully incorporated sulfur chemistry to model soot activation. Soot activation, however, remains a matter of ongoing debate in contrail modeling. While some studies implement explicit soot activation schemes \cite{Wong14, Cantin25}, others consider that soot particles are already activated at the engine exit \cite{Yu24}. Despite their simplified microphysical representations, three-dimensional RANS approaches are particularly valuable as they capture the complete vortex dynamics in the aircraft wake, which strongly influences contrail formation and evolution \cite{UNTER2014}. 

\subsection{Influence of engine position on contrail formation}

At cruise, the engine exhaust is expelled as a propulsive jet carrying engine effluents that initially expands downstream of the aircraft before becoming entrained by the wingtip vortex, wrapping around the vortex structure \cite{GERZ1997}. The latter forms as the vortex sheet on the aircraft wings rolls up. The primary stage of dilution then arises from the initial expansion of the jet near the engine exit. As the flow evolves, the jet becomes increasingly influenced by the vortex, diluting the plume into the atmosphere. This sequence of jet dispersion drives temperature changes due to the entrainment of colder atmospheric air and alters the concentrations of plume species through expansion and mixing with ambient compounds. These evolving thermodynamic conditions and compositions within the plume are key to the chemical and microphysical processes that govern ice crystal formation, which in turn leads to contrail formation \cite{PAOLI2005}. One critical parameter in jet-vortex interactions is the initial distance between the jet and the vortex \cite{GAR1996}. The impact of this distance has been extensively studied through wind tunnel experiments \cite{JAC2007,MAR2008,BOELLE2023} and numerical simulations such as LES \cite{PAOLI2003,LABBE2007}. Adjusting this distance influences vortex dynamics and kinematics, which in turn affect jet properties, including temperature and dispersion. For its part, adjusting the engine's position under the wing impacts jet dilution behind the aircraft. Such adjustments affect contrail properties, which encompass both the spatial distribution and microphysical aspects, such as the ice crystal number, size, and distribution \cite{PAO2013}.

A recent study \cite{PS2023} has provided a more precise analysis of these effects over a longer time. The engine position has been shown to significantly influence the optical properties of contrails, after several minutes of evolution, under different atmospheric stratification conditions. However, the study employed a 2D assumption, initializing calculations in the vortex phase, where the axial jet velocity becomes negligible compared to the transverse velocities of the vortices. As a result, the near-field dynamics of the aircraft are neglected at initialization, omitting critical physical processes such as the influence of aircraft geometry, potential flow separations, compressible effects, the interplay between the high axial velocity of the jet and the rotational vortex flow immediately after ejection, as well as ice crystal formation. Consequently, its impact on contrail formation has not been accounted for. Even more recently, another study has investigated the influence of the engine position on contrail formation and evolution using the RANS approach and a realistic aircraft configuration, incorporating the fuselage, wings, and engines \cite{RR2024}. By simulating the initial jet phase, this work has examined four different engine positions and has demonstrated a clear impact of engine placement on the formation and average properties of the contrail, including ice crystal radius and optical thickness.

\subsection{Objectives of the present study}

Building on these findings, we present an investigation into the effects of engine installation on contrail formation and evolution. Our approach involves conducting RANS simulations to model the flow dynamics around an aircraft configuration while considering the aforementioned near-field effects on contrail formation. To this end, we explore multiple engine positions inspired by real-life configurations and employ a comprehensive aircraft model, incorporating the fuselage, wings, engines, and tailplane—the latter being absent from the Ramsay et al. \cite{RR2024} study and shown to have a significant impact on vortex dynamics \cite{YOU24}. Unlike the aforementioned study, our approach also integrates a mesh adaptation process, enabling accurate resolution of the interaction between the propulsive jet and the wingtip vortex. Additionally, for each engine position, we consider two distinct soot activation scenarios: one involving activation through adsorption processes and the other assuming fully activated soot. In the latter, the sensitivity of contrail formation and evolution to engine position is assessed in comparison to ambient relative humidity and to the soot number emission index. These scenarios enable a more detailed characterization of dilution effects on microphysical processes but also a comparison of engine position sensitivity with other first-order parameters influencing microphysical properties. The impact of engine position on the aircraft's near-field aerodynamics and dilution is analyzed, along with its consequences for microphysical processes and, ultimately, contrail formation.

\section{Methodology}
\label{sec:section2}

\subsection{Fluid flow model}

All numerical studies are carried out using CEDRE, an in-house computational platform developed at ONERA \cite{REFLOCH}. Within this framework, the CHARME solver is specifically employed. CHARME is designed to solve the compressible Navier–Stokes equations for compressible, reactive and multi-species flows. These equations are solved using a cell-centered finite volume method for unstructured meshes, which is particularly well-suited for handling complex geometries, such as those of an aircraft. The RANS equations are used, with variables decomposed into a mean part, $\overline{\Phi}$, and a fluctuation part $\Phi'$, such that $\Phi = \overline{\Phi} + \Phi'$. For compressible flows, an additional density-weighted, time-averaged decomposition (Favre average) is applied, where the variables are expressed as $\tilde{\Phi} = \overline{\rho\Phi} / \overline{\rho}$ and $\overline{\Phi} = \tilde{\Phi} + \Phi''$. The conservation equations implemented in the CEDRE code can then be expressed as follows:

\begin{equation}
\frac{\partial}{\partial t} ( \overline{\rho} \tilde{y}_{k} ) + \frac{\partial}{\partial x_{j}}(\overline{\rho}\tilde{u}_{j}\tilde{y}_{k}) = \frac{\partial}{\partial x_{j}} \left (\overline{\rho}D_{k}\frac{\partial \tilde{y}_{k}}{\partial x_{j}}-\overline{\rho}\widetilde{u^{''}_{j}y^{''}_{k}} \right)   + \overline{\dot{w}}_{k} \label{eq:equation1}
\end{equation}
\begin{equation}
\frac{\partial}{\partial t} ( \overline{\rho} \tilde{u}_{i} ) + \frac{\partial}{\partial x_{j}}(\overline{\rho} \tilde{u}_{j} \tilde{u}_{i}) = -\frac{\partial \overline{p}}{\partial x_{i}} - \overline{\rho}g\delta_{i3} + \frac{\partial}{\partial x_{j}}(\mu \tilde{S}^{d}_{ij} - \overline{\rho}\widetilde{u^{''}_{i}u^{''}_{j}}) \label{eq:equation2}
\end{equation}
\begin{align}
\begin{split}
\frac{\partial}{\partial t} ( \overline{\rho} \tilde{e}_{t} ) + \frac{\partial}{\partial x_{j}}(\overline{\rho}\tilde{u}_{j}\tilde{h}_{t}) & = \frac{\partial}{\partial x_{j}}  \Bigg(\overline{\rho}c_{p}\alpha\frac{\partial \overline{T}}{\partial x_{j}}+\sum_{k}\overline{\rho}\tilde{h_{t}}D_{k}\frac{\partial \tilde{y}_{k}}{\partial x_{j}} - \sum_{k}\overline{\rho}\tilde{h}_{t}\widetilde{u^{''}_{j}y^{''}_{k}} \\ & - \overline{\rho}\widetilde{u^{''}_{j}T^{''}} + 2\mu\tilde{S}^{d}_{ij}\tilde{u}_{i} - \overline{\rho}\widetilde{u^{''}_{i}u^{''}_{j}}\tilde{u}_{i} \Bigg)\label{eq:equation3}
\end{split}
\end{align}

\noindent where $\rho$, $u_i$, $p$, $T$, $y_k$, and $e_t$ denote, respectively, the density, velocity components in the three spatial directions, static pressure, static temperature, species mass fraction, and total energy per unit mass. The Reynolds tensor $\widetilde{u^{''}_iu^{''}_j}$ is modeled using a Boussinesq hypothesis. The two-equations turbulence model $k - \omega$ \textit{SST} (Shear Stress Transport) is used \cite{MEN93}, allowing equation closure. The model relies on transport equations for both the turbulent kinetic energy $k$ and the specific dissipation rate $\omega$, enabling the characterization of small-scale turbulent structures. This turbulence model, with the \textit{SST} correction, is selected as it allows for an accurate representation of near-wall effects and the strong adverse pressure gradients present on a transonic aircraft wing, which would be inadequately captured by a $k-\varepsilon$ model \cite{WIL06}. While it is slightly less precise in resolving rotational flows compared to a Reynolds Stress Model (RSM) \cite{CHURCH09}, it is significantly more computationally efficient. Therefore, this choice represents a balance between accuracy, robustness, and computational cost. The turbulent diffusion fluxes for species species $\widetilde{u^{''}_iy^{''}_k}$ and heat $\widetilde{u^{''}_iT^{''}}$ are expressed by analogy with molecular diffusion fluxes as:

\begin{equation}
    \widetilde{u^{''}_iu^{''}_j} = -\nu_t\tilde{S}^{d}_{ij} + \frac{2}{3}k\delta_{ij}
\end{equation}

\begin{equation}
    \widetilde{u^{''}_iT^{''}} = -\frac{\nu_t}{Pr_t} \frac{\partial \overline{T}}{\partial x_j}
\end{equation}

\begin{equation}
    \widetilde{u^{''}_iy^{''}_k} = -\frac{\nu_t}{Sc^t_k} \frac{\partial \tilde{y}_k}{\partial x_j}
\end{equation}

\noindent where $\nu_t$ represents the turbulent eddy viscosity, $Pr_t$ denotes the turbulent Prandtl number and $Sc^t_k$ corresponds to the turbulent Schmidt number for species $k$.\\

\subsection{Anisotropic mesh adaptation}

To achieve meshes that accurately capture the complex phenomenon within an aircraft wake, we employ an anisotropic mesh adaptation method, using INRIA's feflo.a software \cite{FEFLO}. This technique refines the mesh in areas of interest, based on a specific sensor. It prioritizes refinement in regions where relevant parameter (a user-selected criterion) gradients are significant while maintaining a coarser mesh in areas with weaker gradients \cite{LOSE11,LOSp211}. This approach, validated for its ability to handle aerodynamic complexities stemming from aircraft geometry, has also proven effective in simulating contrails \cite{MON2018}. The procedure involves initially converging to an initial solution $(S_0)$, on an initial mesh $(M_0)$. A criterion is then computed to generate a new mesh, $(M_{i+1})$. The previous solution is interpolated onto the new mesh, $(M_{i+1}, S_{i+1}^0)$, and the convergence process is repeated, gradually increasing the adaptation complexity. The underlying idea is to use this method to obtain an accurate aerodynamic field, enabling to capture of all the intriguing aerodynamic phenomena occurring downstream of the aircraft. A detailed methodology can be found in the reference \cite{ALAU2019}. For the three cases, the chosen sensor is the square of the total kinetic energy in the ground reference frame ($KEG$):

\begin{equation}
\sqrt{KEG} = \sqrt{\frac{1}{2}(U_i-U_{i,\infty})^2 + (k-k_{\infty})} \label{eq:equation7}
\end{equation}

\noindent where $U_i$ are the speed components in the ground frame, $U_{i,\infty}$ are the speed components of the surrounding atmosphere, $k$ represents the mean turbulent kinetic energy in the ground frame and $k_\infty$ denotes the turbulent kinetic energy of the atmosphere at the aircraft's altitude. With these contributions, this criterion allows to efficiently refine both the vortex sheet and the wake vortex, as well as the propulsive jet \cite{YOU2023}.

Mesh adaptation is carried out on the purely aerodynamic field, excluding microphysical processes and considering only the propulsive jet and the airflow around the aircraft. This approach reduces computational cost and avoids the need to transport all chemical species during the adaptation process. Once the aerodynamic field has converged, according to the chosen mesh adaptation metric, it is used as the initial condition for introducing microphysical processes relevant to contrail formation.

\subsection{Microphysics modeling}

Given the low energy contribution of microphysical processes, it is assumed that these do not alter the aerodynamic field. Consequently, microphysical processes can be integrated after the aerodynamic simulations are completed. In the modeling approach used \cite{Khou15, KHOU2016}, which is based on a Eulerian framework, the aircraft plume is initially assumed to consist of gaseous species, treated as perfect gases, along with soot particles and jet temperature. These soot particles, together with the ice crystals that condense onto them, are transported within the flow as passive scalars. This is done using a transport equation applied to the mean particle density:

\begin{equation}
\frac{\partial}{\partial t} ( \overline{\rho} \tilde{N}_{p} ) + \frac{\partial}{\partial x_{j}}(\overline{\rho}\tilde{u}_{j}\tilde{N}_{p}) = \frac{\partial}{\partial x_{j}} \left (\overline{\rho}D\frac{\partial \tilde{N}_{p}}{\partial x_{j}}-\overline{\rho}\widetilde{u^{''}_{J}N^{''}_{p}} \right) \label{eq:equation8}
\end{equation}

\noindent where $N_p$ represents the total particle number density. This quantity is conserved over the volume of the contrail, and particles, whether soot or ice crystals, are distinguished by the presence or absence of condensed ice. In this modeling approach, particles are assumed to be in dynamic and thermal equilibrium within the plume. While this assumption may not hold completely, especially regarding dynamic equilibrium when crystals reach a substantial size, it remains broadly reasonable in the near field given the ice crystal size and their aerodynamic characteristic time \cite{Khou15}. Due to the computational complexity involved in solving the transport equation for each particle size in a polydisperse system, only a single soot particle size of radius $r_s$ is considered in this study. 

As previously mentioned, two soot activation scenarios are considered. The first scenario involves soot particle activation occurring exclusively through the adsorption of sulfuric acid (H\textsubscript{2}SO\textsubscript{4}) and sulfur trioxide (SO\textsubscript{3}) emitted by the engine. These adsorbed species are modeled using species mass fraction, denoted $Y_i^a$. The production of adsorbed species is then governed by a source term $\overline{\dot{\omega}}_{a}^i$ in the mass conservation equation (\Cref{eq:equation1}):

\begin{equation}
\overline{\dot{\omega}}_{a}^i = - \frac{\pi r_s^2 \alpha_{a}(\overline{T}) \rho Y_i \tilde{N}_p \nu_{th,i}(\overline{T})}{M_i}\big(1-\theta_{a}\big),\quad i\in\left\{\text{SO}_3,~\text{H}_2\text{SO}_4\right\}
\label{eq:equation9}
\end{equation}

\noindent where $\nu_{th,i}$ represents the mean thermal velocity, $Y_i$ is the mass fraction, $M_i$ is the molar mass of sulfur species $i$ and $\alpha_a(\overline{T})$ is the accommodation coefficient for the sulfur species. This coefficient is set to 1 if $T\leq420K$ and 0 otherwise. The surface fraction of activated soot, $\theta_a$, is then determined based on the concentrations of the adsorbed species \cite{WONG08}:

\begin{equation}
    \theta_{a} = \frac{N_A}{4\pi r_{s}^2\tilde{N}_p \sigma_0}([\text{SO}_{3}^{a}]+[\text{H}_2\text{SO}_4^{a}]) \label{eq:equation10}
\end{equation}

\noindent where $N_A$ represents the Avogadro's number, $\sigma_0$ denotes the number of available adsorption sites per unit area of a soot particle \cite{KAR1998} and $[\text{SO}_{3}^{a}]$ and $[\text{H}_2\text{SO}_4^{a}]$ are the concentrations of sulfur species adsorbed onto the soot surface.

Once the soot particles are activated, water vapor condensation can initiate at these sites. The model assumes that nucleation occurs exclusively through heterogeneous processes and that ice growth is restricted to the surfaces of activated soot particles. Crystal growth is then scaled according to the fraction of activated soot. Water vapor condenses onto the soot particles, with their growth governed by Fick's law. Ice is modeled through the mass fraction $Y_{H_2O,s}$, while phase transitions (condensation and evaporation) are modeled by the reversible reaction: $H_2O,v \leftrightarrow H_2O,s$. The mass transfer rate is included as a source term in the mass conservation equation (\Cref{eq:equation1}) within the RANS framework:

\begin{equation}
    \overline{\dot{\omega}}_{ice}=\frac{4\pi \tilde{N}_p r_p \theta_{a}D_{H_2O,v}M_{H_2O}}{R\overline{T}}\bigg(p_{H_2O,v}-p_{H_2O}^{sat,ice}(\overline{T})\bigg)G(r_p)\Pi(p_{H_2O}^{sat,liq}(\overline{T}),r_p) \label{eq:equation11}
\end{equation}

\noindent where $r_p$ is the particle radius, $D_{H_2O,v}$ is water vapor diffusion coefficient, and $p_{H_2O,v}$, $p_{H_2O}^{sat, ice}$, $p_{H_2O}^{sat,liq}$  represent, respectively, the partial pressure of water vapor, the saturation pressure with respect to ice, and the saturation pressure with respect to liquid water. In this equation, the $G(r_p)$ function accounts for the transition between the kinetic gas regime and the diffusion regime. Its formulation is based on the Knudsen number $Kn$ and follows the approach proposed by the reference \cite{DAVIES76}:

\begin{equation}
    G(r_p) = \bigg(\frac{1}{1+Kn(r_p)}+\frac{4Kn(r_p)}{3\alpha}\bigg)^{-1}
\end{equation}

\noindent where $\alpha$ represents the deposition coefficient of water molecules onto ice, set to 0.1 \cite{KAR96}. The function $\Pi(p_{H_2O,v}^{sat,liq},r_p)$  triggers condensation on particles saturated with liquid water, a critical condition for the formation of ice crystals in contrails \cite{JEN98}:

\begin{equation}
    \Pi(p_{H_2O}^{sat,liq}(T),r_p) = 
    \begin{cases}
        0 & \text{if } p_{H_2O,v} \le p_{H_2O}^{sat,liq}(T) \text{ and } r_p = r_s \\
        1 & \text{if } p_{H_2O,v} \ge p_{H_2O}^{sat,liq}(T) \text{ or } r_p > r_s
    \end{cases}
    \label{eq:equation13}
\end{equation}

The second activation scenario assumes that soot particles are fully activated. In this case, sulfur species are excluded from the calculation, and the surface activation fraction is set to $\theta_a = 1$ (\Cref{eq:equation11}). As a result, particle growth is no longer weighted by the surface area of activated soot. This assumption implies that the saturation ratio, expressed as:

\begin{equation}
    S_K = \text{exp}\bigg(\frac{2\sigma M_{H_2O}}{R\overline{T}\rho_vr_p}\bigg) \label{eq:sk}
\end{equation}

\noindent is equivalent to applying $\kappa$-Köhler theory \cite{PET07} with the hygroscopicity parameter set at $\kappa$=0, and corresponding to pure hydrophobic soot.

Finally, the primary distinction of the model in the present work from the previous model \cite{KHOU2016} lies in the treatment of chemical reactions. In this study, significant reductions in computation time are achieved by simplifying the reaction scheme. The original scheme, which comprised 23 species, is reduced to 7 key species for cases involving soot activation by sulfur species: water vapor (H\textsubscript{2}O), the sulfur compounds responsible for soot activation (H\textsubscript{2}SO\textsubscript{4} and SO\textsubscript{3}), the adsorbed species ($\text{H}_2\text{SO}_4^a$ and $\text{SO}_3^a$), the ice ($\text{H}_2\text{O},s$) and the main constituents of ambient air, modeled using an equivalent mass fraction $Y_{air}$, which does not interact with the other species. The model considers only three species for fully activated scenarios: water vapor, ice, and the equivalent species representing air. This simplification, for modeling cases with sulfur species, is made based on the observation that the vast majority of chemical reactions are mainly completed just before the onset of microphysical processes. This conclusion is further supported by the trends that are observed in reference \cite{KHOU2016}.

\section{Numerical setup}
\label{sec:section3}

The aircraft configuration used for the calculations is the Common Research Model (CRM), a geometry developed by NASA to represent a Boeing 777 \cite{CRM}. This configuration has been adapted by ONERA to include two engines, each representative of engines with a bypass ratio equal to 12 \cite{MON2018}, enabling the simulation of the propulsive jet with both core and bypass flows. The complete geometry includes the fuselage, wings, engines, and tailplane, the latter of which was not considered in the study by Ramsay et al. \cite{RR2024}. All CRM characteristics, engine inlet and outlet conditions, and atmospheric parameters used in each simulation are summarized in \Cref{tab:table1} and detailed in the following paragraphs.

\begin{table}[H]
    \centering
    \caption{Simulation parameters: aircraft geometry, engine specifications, exhaust composition, and flight conditions}
    \label{tab:table1}
    \renewcommand{\arraystretch}{1.3}
    \begin{tabular}{c}
    \toprule

        \textbf{CRM Dimension} \\
        \midrule
        \begin{tabular}{c c c c c c}
            Wingspan & $b$ & 58.8 m &
            Fuselage length & $L$ & 62 m \\
            Wing area & $S$ & 383 m$^2$ &
            Mean chord & $c_m$ & 6.5 m \\
        \end{tabular} \\
        
        \midrule
        
        \textbf{Engine thermodynamic conditions} \\
        \midrule
        \begin{tabular}{c c c c c c}
            \multicolumn{6}{c}{\textbf{Inlet}} \\
             & Total pressure & $P_{t,in}$ & 370.12 hPa & & \\
            \cmidrule(lr){1-6}

            \multicolumn{6}{c}{\textbf{Core}} \\
            Total temperature & $T_{t,c}$ & 626.41 K & Total pressure & $P_{t,c}$ & 530.0 hPa \\
            Outlet velocity & $U_{c}$ & 458 m/s & Outlet area & $A_c$ & 0.53 m$^2$ \\
            
            \cmidrule(lr){1-6}
            
            \multicolumn{6}{c}{\textbf{Bypass}} \\
            Total temperature & $T_{t,bp}$ & 297.23 K & Total pressure & $P_{t,bp}$ & 700.0 hPa \\
            Outlet velocity & $U_{bp}$ & 310 m/s & Outlet area & $A_{bp}$  & 3.30 m$^2$ \\
        \end{tabular} \\

        \midrule
        \textbf{Chemical exhaust composition (mass fractions $Y_k$)} \\
        \midrule
        \begin{tabular}{c c c c}
            \ch{H2SO4} & $1.4 \times 10^{-8}$ &
            \ch{H2SO4^a} & 0.0 \\
            \ch{SO3} & $3.0 \times 10^{-10}$ &
            \ch{SO3^a} & 0.0 \\
            \ch{H2O} & 0.0206 &
            \ch{Air} & $1 - \sum Y_k$ \\
        \end{tabular} \\
        
        \midrule
        
        \textbf{Soot parameters} \\
        \midrule
        \begin{tabular}{c c c c c c}
            Soot number emission index & EI\textsubscript{soot} & $1 \times 10^{14}$; $1 \times 10^{15}$ $\text{kg-fuel}^{-1}$ &
            Radius & $r_s$ & 27 nm \\
        \end{tabular} \\
        
        \midrule
        
        \textbf{Flight conditions and domain dimensions} \\
        \midrule
        \begin{tabular}{c c c c c c}
            Temperature & $T_\infty$ & 223.15 K &
            Pressure & $P_\infty$ & 264.37 hPa \\
            Humidity (ice) & $RH_i$ & 110\%; 120\% &
            Mach number & $M_\infty$ & 0.85 \\ 
            Velocity & $U_\infty$ & 254.38 m/s &
            Angle of attack & $\alpha$ & $3.0^\circ$ \\
            \multicolumn{6}{c}{Domain dimensions $\quad L_x\,$, $L_y\,$, $L_z$ $\quad32b\,$, $10b\,$, $20b$} \\
        \end{tabular} \\

        \bottomrule
    \end{tabular}
\end{table}

\begin{figure}[!ht]
\centering
\includegraphics[width=0.9\linewidth]{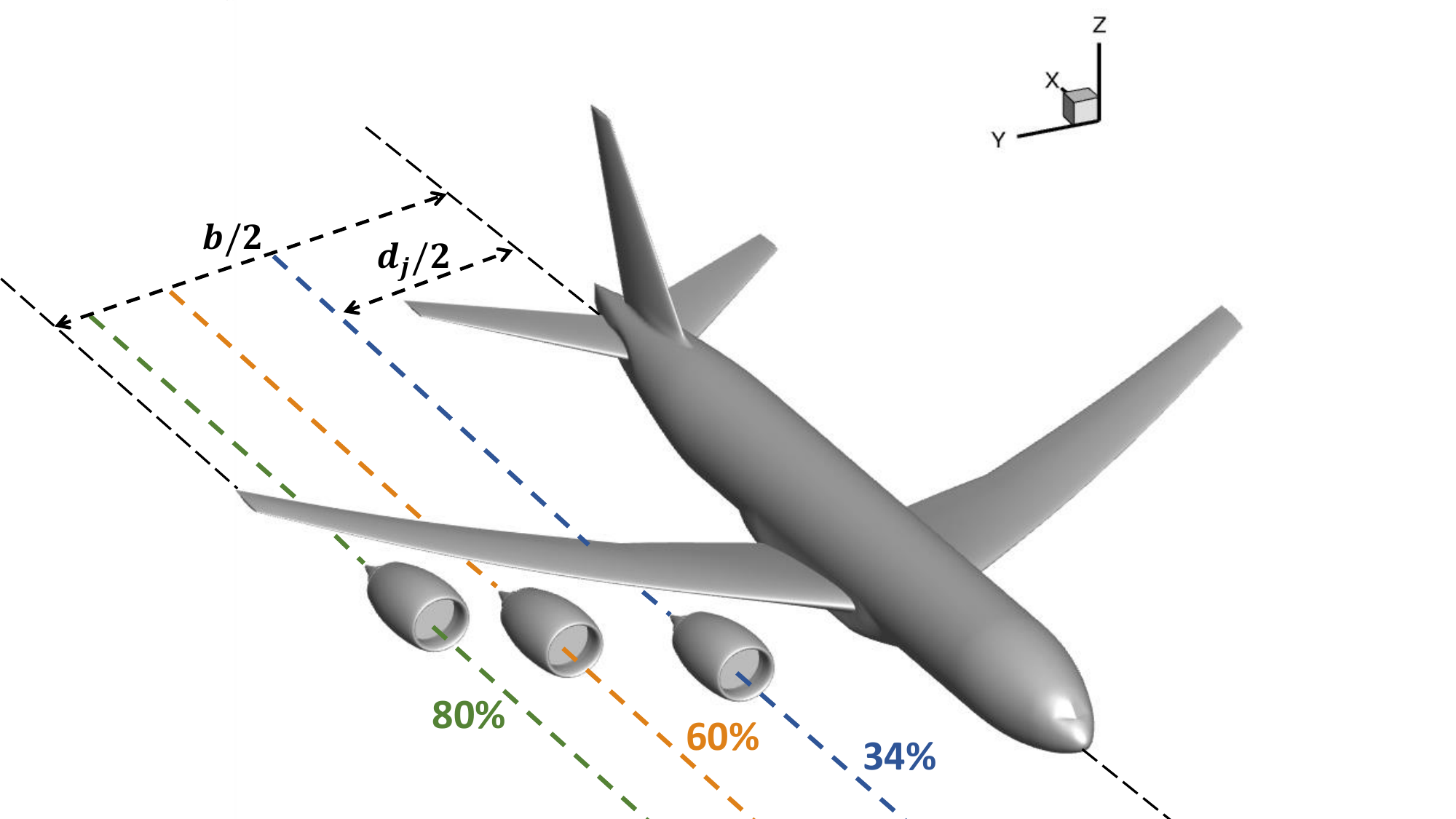}
\caption{CRM simulation configurations: $d_j / b=$ 0.34 (blue); 0.60 (orange); 0.80 (green)}
\label{fig:figure1}
\end{figure}

Three configurations, each with distinct engine positions, are studied. The first configuration corresponds to the original CRM engine position, at $d_j/b$ = 0.34, meaning the engines are positioned at 34\% of the half-wing span (\Cref{fig:figure1}), where $d_j$ is the distance between the jets. This setup approximates the engine placement of aircraft such as the A-320, A-350, or B-777. The second configuration places the engines at $d_j/b$ = 0.60, or 60\% of half-wing span, representing the outboard engine position on a B-747 or near to that of an A-380. The third configuration corresponds to $d_j/b$ = 0.80, aligning the jet with the final horizontal position of the wingtip vortices, which are separated by a distance $b_0 = \frac{\pi}{4}b$ under elliptic loading conditions \cite{JAC2001}. This places the jet at approximately 80\% of the wing’s semi-span. This position has also been identified as having a notable impact on the optical properties of contrails after several minutes of evolution, notably reducing extinction compared to other engine placements \cite{PS2023}. In all configurations, the pylon connecting the engine to the wing is removed to allow for engine repositioning along the wing. 

The symmetry is leveraged to perform simulations for only one-half of the aircraft. The computational domain is defined as a rectangular prism extending over 22 wingspans downstream of the aircraft, equivalent to approximately 1.3 km. The aircraft is positioned 10 wingspans from the domain entrance and equidistant from the upper and lower boundaries, resulting in domain dimensions of $L_x=32b$, $L_y =10b$, and $L_z=20b$. This configuration ensures that the boundary conditions, representing the surrounding atmosphere, have minimal influence on the flow near the aircraft. The thermodynamic data for the engine boundary conditions are obtained from the reference \cite{MON2018}, adapted from a CFM56-3 engine \cite{GAR1997} to be integrated into the initial CRM nacelle, while being more representative of a modern high-bypass-ratio engine, with a bypass ratio set to 12. Based on the provided data and boundary conditions, the engine specifications were derived. The equivalent engine would have a fuel consumption of 0.67 kg/s and an overall efficiency of 0.33.

For each engine position, two scenarios are considered: one involving soot activation through adsorption and another assuming fully activated soot. These activation scenarios make it possible to distinguish the specific effects of dilution on contrail formation. In the sulfur-driven activation case, dilution affects the activating species, thereby isolating its influence on the activation process. Conversely, the fully activated soot case allows a focus solely on the effects of dilution on water vapor and soot. The mass fractions, originally derived from a CFM56-3 engine \cite{GAR1997}, are adjusted to align with the new modeling approach, where chemical reactions are not considered. While the water mass fraction remains unchanged, the sulfur mass fractions are modified for the adsorption cases. These adjustments are based on previous simulations conducted with Jet A-1 containing 100 ppm of sulfur. For fully activated cases, only air and water vapor species are considered at the engine's primary outlet, resulting in a higher air fraction than in scenarios with activation through adsorption. A soot number concentration of $N_{p,exit}$ = $3.42\times10^{12}$ $/m^3$, corresponding to an emission index $\text{EI}_{\text{soot}}=1\times 10^{15}\text{ kg-fuel}^{-1}$, is taken as the reference in the present study and serves as a baseline for comparing different activation scenarios. The impact of engine position on contrail properties is compared to the sensitivity to the soot number emission index, an essential parameter in contrail formation \cite{Kar09}. A reduced soot number density of $N_{p,exit}$ = $3.42\times10^{11}$ $/m^3$ (i.e., ten times lower than the reference case) is then employed. This allows for direct comparison with the fully activated scenario at the original soot number concentration. These two emission indices correspond, respectively, to typical values for Jet A-1 and Sustainable Aviation Fuel (SAF) \cite{Markl24}. The soot particle radius size is $r_s$ = 27 nm.

The boundary conditions are based on parameters representative of a cruise flight at an altitude of 10 km. The atmospheric temperature is set to 223.15 K, the atmospheric pressure to 264.37 hPa, and the relative humidity with respect to ice ($RH_i$) to 110\%. This latter value serves as the reference case for the calculations. The sensitivity of contrail formation and evolution to engine position is also evaluated in comparison to that associated with ambient relative humidity, by considering a scenario with 120\% relative humidity. The flight Mach number is set to 0.85, corresponding to a cruise speed of $U_\infty$ = 254.38 m/s. An angle of attack ($\alpha$) of $3.0^\circ$ is chosen.

\section{Aerodynamic results}
\label{sec:section4}

\subsection{Mesh convergence analysis}

\begin{figure}[H]
    \centering
    \includegraphics[width=0.9\textwidth]{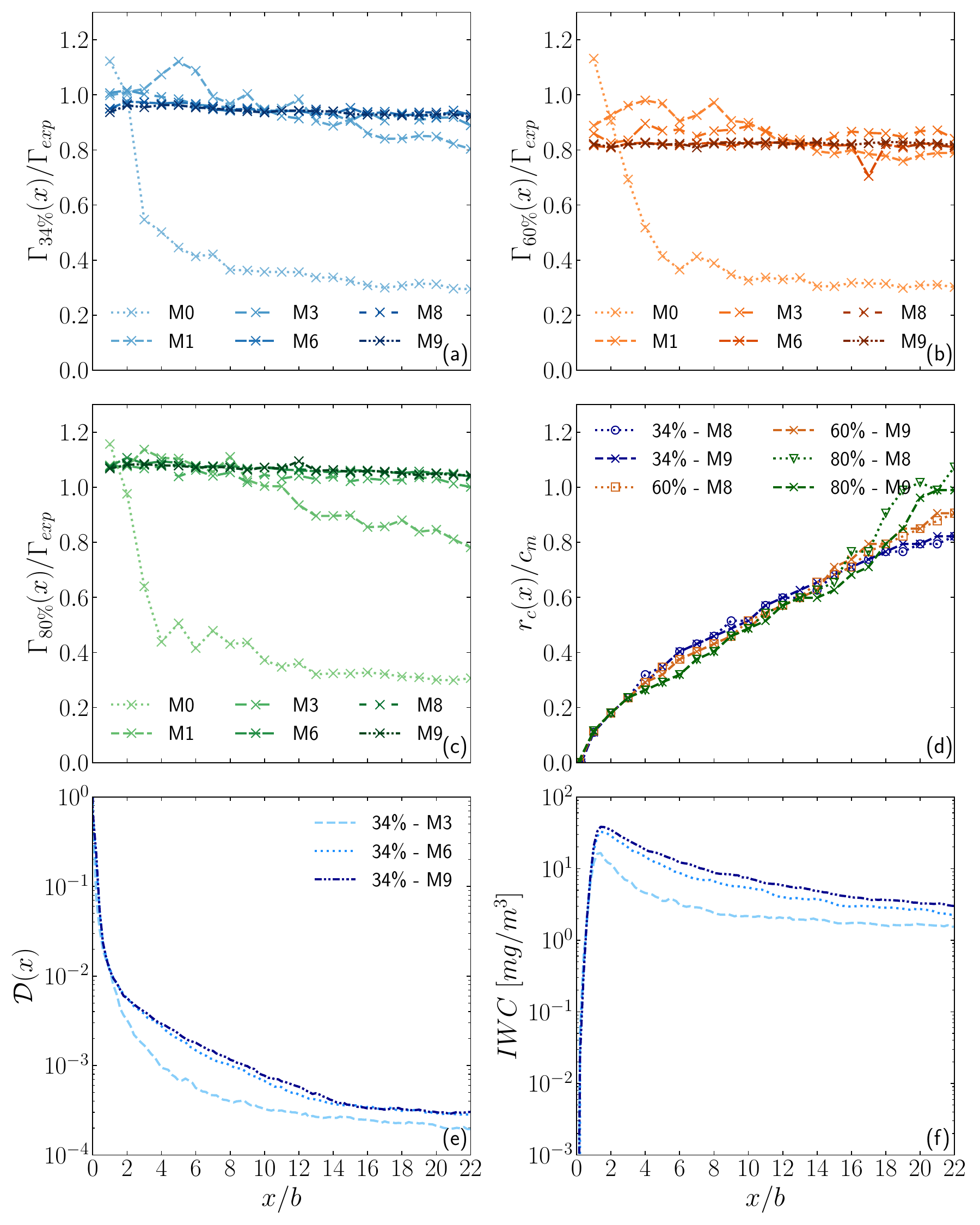}
    \caption{Evolution of the total circulation $\Gamma(x)$ for the three configurations: 34\% (a), 60\% (b), 80\% (c) across meshes iterations (Mi). Vortex radius between mesh iteration n°8 (M8) and n°9 (M9) for the three cases (d). Evolution of (e) plume dilution $\mathcal{D}(x)$ and (f) mean ice water content $IWC$, for meshes M3, M6 and M9, in the case 34\% and $\text{EI}_{\text{soot}}=1\times 10^{15}\text{ kg-fuel}^{-1}$ as a function of the distance behind the aircraft}
    \label{fig:figure2}
\end{figure}

After multiple iterations, the mesh adaptation process has produced a converged aerodynamic field. To ensure that the resulting meshes accurately capture the aircraft's aerodynamic wake for all three configurations, key wake characteristics are evaluated. To this end, the total circulation $\Gamma(x)$ and the vortex radius $r_c(x)$, which are two key parameters characterizing a vortex, are calculated at each iteration of the mesh adaptation process. The total circulation is determined by integrating the vorticity over each $x$-plane behind the aircraft. The vortex radius is measured as the distance from the vortex center to the location of maximum tangential velocity. Both quantities are plotted for the three configurations in \Cref{fig:figure2} in panels (a-d). The circulation obtained from the simulations is normalized using the experimental circulation derived from the reference \cite{RIV11}, where wind tunnel tests determined the lift coefficient for the CRM. The experimental circulation is calculated based on elliptical wing theory, $\Gamma_{exp} = \frac{C_L U_\infty S}{\pi b}$, yielding $\Gamma_{exp} =529~\text{m}^2/\text{s}$.

Likewise, to demonstrate the impact of mesh resolution on contrail formation, simulations incorporating the fully activated microphysics model are conducted using meshes M3, M6, and M9 for the 34\% engine configuration. The dilution factor $\mathcal{D}(x)$ and ice water content ($IWC$, defined as $IWC=\rho Y_{H_2O,s}$) are shown in panels (e) and (f), respectively. The dilution factor is defined by the following expression:

\begin{equation}
    \mathcal{D}(x) = \frac{\overline{N}_p(x)-N_{p,\infty}}{N_{p,exit}-N_{p,\infty}}
    \label{eq:equation15}
\end{equation}

\noindent where $N_{p,\infty}$ represents the soot number particle density in the surrounding atmosphere, equal here to 0. We chose to use the mean particle number density, $\overline{N}_p$, to calculate the dilution factor, as soot particles are treated in our model as passive tracers that do not interact with the flow. Based on this definition, the dilution factor is highest at the engine outlet and decreases progressively as the plume mixes with the surrounding atmosphere.

For all three configurations, the circulation behind the aircraft does not remain consistent for initial mesh adaptation iterations, particularly after a few wingspans. However, after several iterations of mesh refinement, the circulation stabilizes and converges to a value close to the theoretical circulation. The discrepancies between simulation results and experimental values can be attributed to the absence of a propulsive jet in the experiment, which influences the total circulation, as well as differences introduced by engine relocation. Additionally, the use of a turbulence model may have contributed to the dissipation of rotational effects, resulting in some losses in total circulation \cite{CHURCH09}. In the final mesh, the circulation is well-preserved, with only minimal differences between the last and second-to-last iterations. For the vortex radii of each configuration, good agreements are observed between the final iterations, with radii converging to consistent values. In the 80\% configuration, a slight divergence occurs beyond 15 wingspans due to the difficulty of measuring the vortex radius as it begins to merge with another vortex (discussed further in the results). Nevertheless, the trends align well with wind-tunnel observations, which indicates that the jet has no significant impact on the vortex size \cite{JAC2007,BOELLE2023}. This consistency in circulation and vortex radius across iterations demonstrates the mesh’s ability to accurately capture the key aerodynamic phenomena in the wake.

Mesh resolution has a significant impact on microphysical processes. As shown in panel (e), dilution occurs significantly more rapidly with the M3 mesh, the coarsest resolution, compared to M6 and M9. This behavior is primarily due to excessive numerical diffusion in the under-resolved M3 plume, which leads to artificial dissipation and dispersion of soot particles in the wake. From an aerodynamic standpoint, this corresponds to poorly conserved circulation, resulting in an overly diffused vortex with an exaggerated radius that amplifies plume dispersion. In contrast, the M6 and M9 meshes yield much more consistent dilution values, although M6 still overestimates dilution between 4 and 14 wingspans behind the aircraft. The overestimation of plume dilution in the atmosphere is directly reflected in the amount of ice formed within the plume (panel (f)). The ice content of each contrail decreases with coarser mesh resolution, primarily due to the inability to accurately resolve temperature and species gradients, both of which are critical for modeling condensation. This under-resolution not only affects particle dilution and ice mass but also impacts key microphysical properties such as ice crystal radius and contrail optical thickness. These results underscore the importance of using an aerodynamically converged mesh, which enables a more accurate representation of turbulent mixing in the aircraft wake and the resulting formation of induced contrails.

\subsection{Impact on jet dilution}

\begin{figure}[ht]
\begin{center}
\includegraphics[width=1.\textwidth]{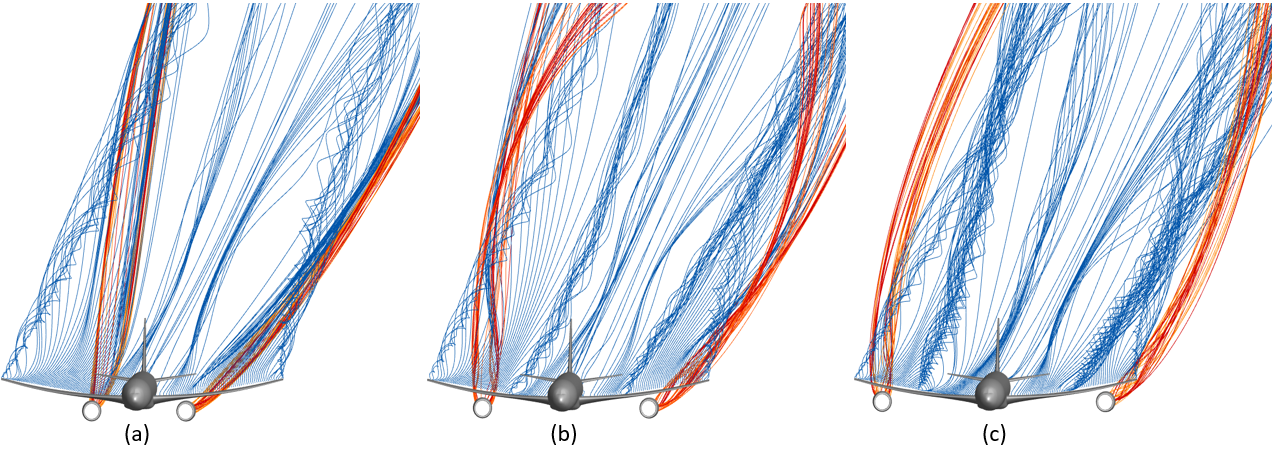}%
\end{center}
\caption{Streamlines colored by their originating element: wings (blue), primary flow from the engine (red), and secondary flow from the engine (orange) for the three configurations: (a) 34\%, (b) 60\% and (c) 80\%}
     \label{fig:figure3}
\end{figure}

Changes in engine position significantly impact the vortex dynamics and kinematics, as reflected in the topology of the downstream streamlines for each configuration (\Cref{fig:figure3}). 

These variations in jet-vortex interactions, particularly in the velocity fields near the engine, lead to notable differences in jet dispersion among the configurations. In the 34\% configuration, the jet is gradually wrapped around the vortex; in the 60\% configuration, it is partially entrained within the vortex; and in the 80\% configuration, it remains relatively compact and localized around the vortex. In the latter two cases, additional vortex structures form, which is not the case in the 34\% configuration. The nacelle's position influences the formation of these structures, which subsequently alter the vortex dynamics, either joining the tail wake in the 60\% case or merging with the wingtip vortex in the 80\% case. It is important to note that the vortex structures observed here differ in their dynamics from those described by Ramsay et al. \cite{RR2024}. In their study, engine displacement configurations excluded the tailplane, which significantly affects the vortex dynamics in the present study due to the influence of the tail vortex.
Furthermore, the CRM wing, which is already prone to flow separation under certain flight configurations \cite{BALAK11}, displays additional separation phenomena with changes in engine position. These separation events are driven by the interaction between shock waves generated on the nacelle and those forming on the wing, as well as by a local acceleration of the flow, which intensifies the shock on the wing. These separation phenomena and newly formed vortex structures are particularly noteworthy, as they arise solely from nacelle installation effects. They highlight the critical role of 3D simulations in uncovering complex interactions and dynamics that would otherwise remain undetected in simpler modeling approaches.

\begin{figure}[ht]
\begin{center}
\includegraphics[width=.5\textwidth]{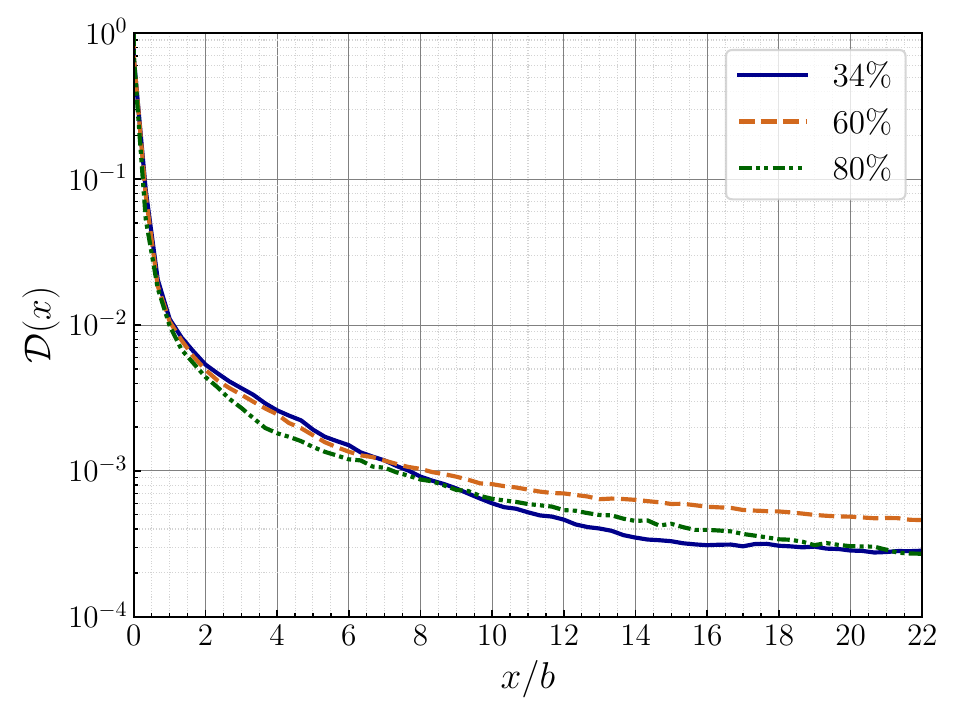}%
\end{center}
\vspace{-0.6cm}
\caption{Evolution of plume dilution $\mathcal{D}(x)$ for the three configurations: 34\% (solid line), 60\% (dashed line), and 80\% (dash-dotted line) as a function of the distance behind the aircraft}
\label{fig:figure4}
\end{figure}

These variations in how the jet interacts with the wingtip vortex lead to changes in the dilution of the plume. This is quantified by plotting the dilution factor $\mathcal{D}(x)$ as a function of the distance downstream of the aircraft (\Cref{fig:figure4}).

\begin{figure}[ht]
\begin{center}
\includegraphics[width=.5\textwidth]{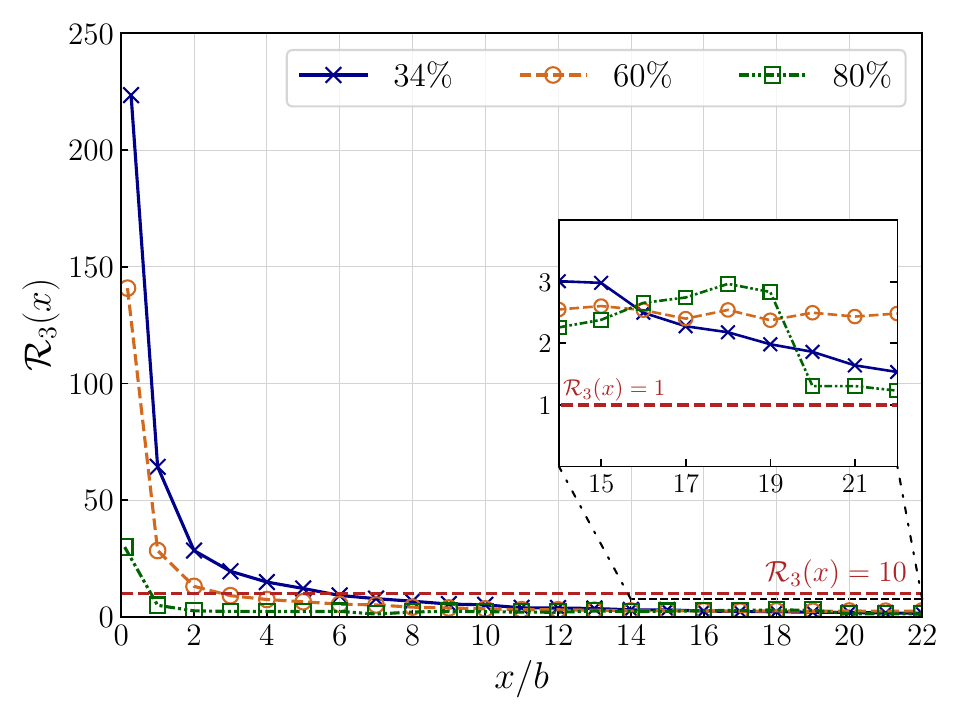}%
\end{center}
\vspace{-0.6cm}
\caption{Evolution of $\mathcal{R}_3(x)$ criterion for the three configurations: 34\% (solid line), 60\% (dashed line), and 80\% (dash-dotted line) as a function of the distance behind the aircraft.}
\label{fig:figure5}
\end{figure}

The graph in \Cref{fig:figure4} can be divided into two distinct regions. The first region extends from the engine outlet to 8 wingspans downstream of the aircraft. In this zone, the 60\% and 80\% configurations show the highest plume dilution, primarily due to the proximity of the propulsive jet to the wingtip vortex, which promotes faster particle dispersion. In contrast, the jet in the 34\% configuration primarily exhibits its own dynamics before the vortex influence becomes substantial. Upon exiting the engine, the jet, dominated by a high axial velocity, is initially unaffected by the transverse velocities of the vortex. However, positioning the engine closer to the wingtip accelerates the jet's deceleration, creating competition between the axial and transverse velocities. This effect can be quantified using the $\mathcal{R}_3(x)$ criterion \cite{GAR1996}, as plotted in \Cref{fig:figure5}, which compares the moments of the jet velocity and the transverse velocity:

\begin{equation}
\mathcal{R}_3(x) = \frac{\rho_j U_j(U_j - U_{\infty})A_j}{\rho_0U_{\theta}^{2}(x)A(x)}
\label{eq:equation16}
\end{equation}

\noindent where $\rho_j$ , $U_j$ and $A_j$ are the jet density, velocity, and area, respectively, while $U_\theta$, $\rho_0$ and $A(x)$ correspond to the aircraft velocity, atmospheric density, and jet area at a downstream location $x$. 

This ratio indicates the transition of the jet from its own dynamics ($\mathcal{R}_3(x)\gg1$) to being influenced by the vortex ($\mathcal{R}_3(x)\ll1$). The two quantities defining the criterion become comparable in magnitude after approximately one wingspan for the 80\% configuration, three wingspans for the 60\% configuration, and seven wingspans for the 34\% configuration. This shift significantly affects the plume’s properties, influencing both its dispersion and the thermodynamic conditions within it. In the second part of the graph, where the jets are fully influenced by the vortex, the flow topology explains the observed trends. In the 34\% configuration, the jet wraps around the vortex, leading to faster particle dilution. In the 80\% configuration, some particles are entrained within the vortex dipole, enhancing dilution as compared to the 60\% configuration, where the jet penetrates the vortex and disperses within it.

The aerodynamics behind each configuration exhibit notable differences, particularly in the interaction between the jet and the vortex (modifying the $\mathcal{R}_3$ value) as well as in jet dilution. These differences are highlighted through mesh adaptation, which optimizes the grid for the selected metric, allowing for computational efficiency gains—both in mesh generation and in calculations that would otherwise suffer from an under- or over-refined grid. Furthermore, the inclusion of the aircraft’s tail in this study is crucial, as it reveals different vortex dynamics induced by flow separation on the wing, which is itself influenced by the displacement of the nacelle. Considering these effects is essential, as the flow field no longer consists solely of a close interaction between the jet and a single vortex; instead, additional near-field aerodynamic effects emerge that could not have been predicted otherwise. As a result, the distinct aerodynamic characteristics behind each configuration, particularly in the way the jet disperses, directly impact the formation and mean properties of contrails associated with different engine positions.

\section{Contrails microphysics}
\label{sec:section5}

\subsection{Aerodynamic influence on contrail expansion}

\begin{figure}[ht]
\begin{center}
\includegraphics[width=0.9\textwidth]{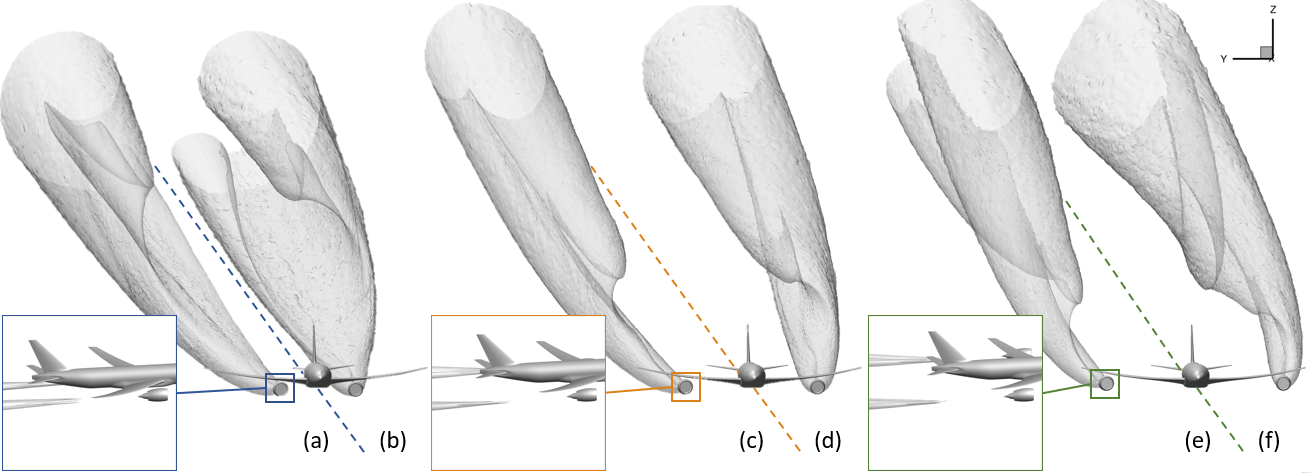}%
\end{center}
\vspace{-0.6cm}
\caption{Visualization of contrails using isosurfaces of Ice Water Content ($IWC = 10^{-6}~mg/m^3$): (a) 34\% activation through adsorption, (b) 34\% fully activated, (c) 60\% activation through adsorption, (d) 60\% fully activated, (e) 80\% activation through adsorption, and (f) 80\% fully activated}
\label{fig:figure6}
\end{figure}

The contrails from the simulations are visualized by plotting the Ice Water Content ($IWC=\rho Y_{H_2O,s}$) isosurface (\Cref{fig:figure6}). For each engine configuration, the left panels show the cases with soot activation through adsorption, while the right panels depict the fully activated cases. The left panels also show a zoomed-in view centered on the engine for each configuration, focusing on the contrail formation region and offering clearer insight into the initial formation processes. The differences observed for the same engine position between the two soot activation methods are primarily due to the visualization criterion used for the contrail. In the fully activated scenario, crystal growth is not weighted, resulting in higher ice content and a thicker appearance of the contrail. Additionally, in these cases, contrails form earlier, as condensation begins as soon as the plume reaches supersaturation with respect to liquid water.

For each engine position, the contrails exhibit distinct shapes, emphasizing the aerodynamic variations induced by differences in the engine placement (\Cref{fig:figure3}). In the 34\% configurations, the contrail progressively wraps around the wingtip vortex throughout the computational domain. By the domain's end, the contrail begins to penetrate the vortex. In the 60\% configurations, part of the contrail enters the vortex, producing a distinctive structure with ice crystals distributed between the jet and the vortex. Finally, in the 80\% configuration, the contrail initially remains compact but eventually spreads into both vortices (the wingtip vortex and the vortex generated by flow separation) as they merge. This results in a layered structure, with one region associated with the jet and another integrating into the merging vortices.

\begin{figure}[ht]
\begin{center}
\includegraphics[width=1.\textwidth]{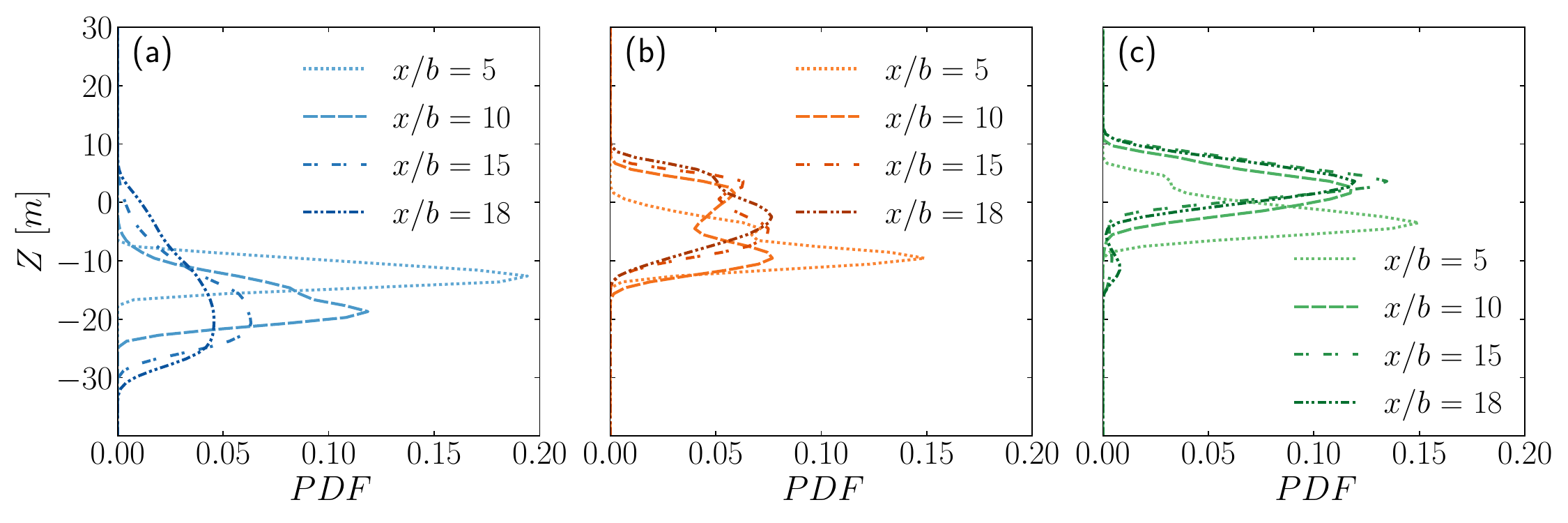}%
\end{center}
\vspace{-0.6cm}
\caption{Evolution of the PDFs of particle altitude for the three configurations: (a) 34\%; (b) 60\%; (c) 80\% at distances $x/b$=5, 10, 15, 18 behind the plane}
\label{fig:figure7}
\end{figure}

The aerodynamic analysis of the three configurations revealed distinct dynamic behaviors, particularly in the descent of the wingtip vortex and the spatial dispersion of the jet. The jet’s entrainment dynamics varied significantly across the configurations, resulting in distinct trajectories around the wingtip vortex and contributing to differences in the spatial spread of the plume (\Cref{fig:figure3}). To quantify these variations, probability density functions (PDFs) are plotted for each engine configuration (different activation cases within the same configuration exhibit similar dynamics), representing the vertical distribution of the plume by showing the altitude as a function of the particle count (\Cref{fig:figure7}).

The particle distribution trends vary significantly across the configurations. In 34\% cases, the plume evolution is gradual, spreading in altitude as it moves downstream from the aircraft. This results in a progressive dispersion pattern, with higher particle concentrations centralized around the jet core, consistent with the overall contrail structure. In contrast, the 60\% and 80\% configurations exhibit distinct behaviors. In the 60\% cases, the plume remains at a higher altitude and displays a dual-peak distribution in the PDF. These two peaks correspond to particles within the jet core and those entrained in the vortex. Meanwhile, the 80\% configuration shows a predominantly high-altitude plume, where most particles are concentrated in the jet core, forming the main peak in the PDF. A smaller secondary peak corresponds to particles entrained in the vortex, representing a reduced yet distinct population.

\begin{figure}[H]
\begin{center}
\includegraphics[width=0.65\textwidth]{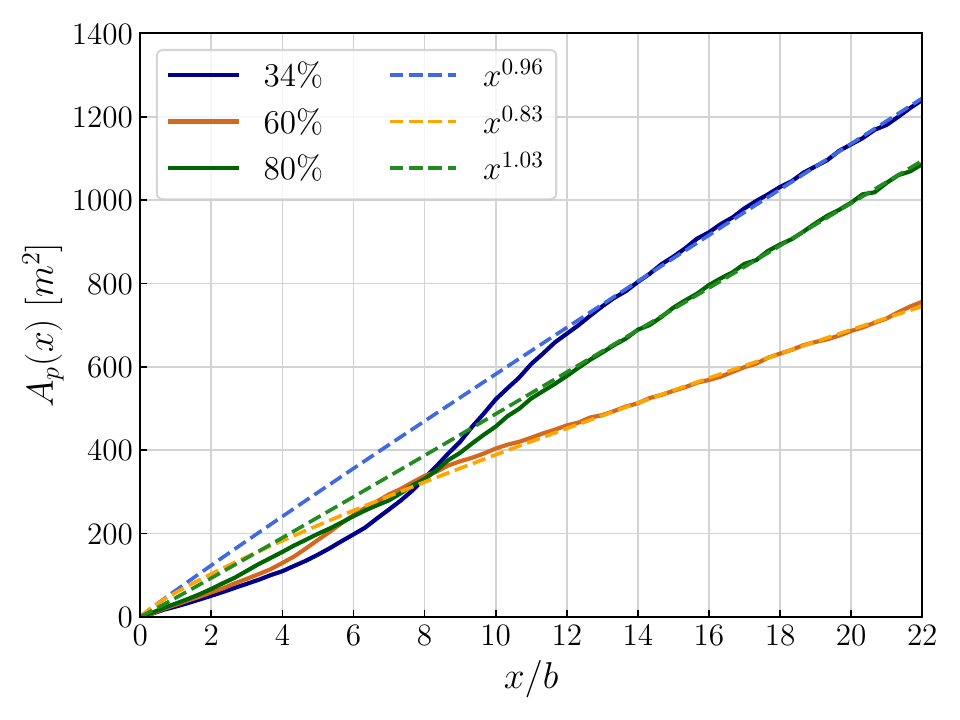}%
\end{center}
\vspace{-0.6cm}
\caption{Evolution of the transverse plume area $A_p(x)$ (solid lines) and the fitted power law (dashed lines) for the three configurations, as a function of the distance behind the aircraft}
\label{fig:figure8}
\end{figure}

This spatial evolution is also reflected in the transverse surface area of the plume (\Cref{fig:figure8}), which expands as the plume spreads. The surface area is approximated using an $\eta_p$ power law (\Cref{eq:equation17}) as in reference \cite{PAO2013}. Here, we plot the cross-sectional areas for different engine positions, considering regions where $N_p > 1~m^{-3}$. As a result, the cross-sectional areas remain the same between cases with different activation methods.

\begin{equation}
    A_p(x) = A_p(x_0)\bigg(\frac{x}{x_0}\bigg)^{\eta_p}
    \label{eq:equation17}
\end{equation}

This power law is commonly used for initializing Gaussian plumes in simplified simulations, and serves as an indicator of dilution dynamics derived from in-flight measurements \cite{SCHU98}. The obtained exponent values align closely with those observed for plumes from various aircraft, reinforcing the validity of the dilution dynamics in the simulation. The reference distance, $x_0$, is set at 16 wingspans to exclude the jet entrainment phase during the initial moments, which is non-representative of the plume’s longer-term behavior. In terms of order of magnitude, the values are consistent with those reported in the literature \cite{SUSS01,PAO2013,KOLO18}. For the 34\% configuration, the plume undergoes the most significant evolution. Initially, the jet experiences a gradual entrainment phase, resulting in a narrower plume as compared to the other configurations. This reduced initial width is consistent with the findings of Paoli et al. \cite{PAOLI2003}, which show that jets farther from the vortex undergo prolonged interaction phases, leading to narrower plumes during the early stages as compared to configurations where the jet is closer to the vortex. In the 80\% configuration, the plume exhibits a larger surface area as compared to the 60\% configuration, primarily due to particles entrained into the vortex dipole. Although the particle density in this region is relatively low, these particles contribute to the overall plume surface area, as they are accounted for in the calculation of $A_p(x)$.

\subsection{Effect of dilution on microphysics properties}

\subsubsection{Impact on plume saturation}

\begin{figure}[ht]
\begin{center}
\includegraphics[width=1.\textwidth]{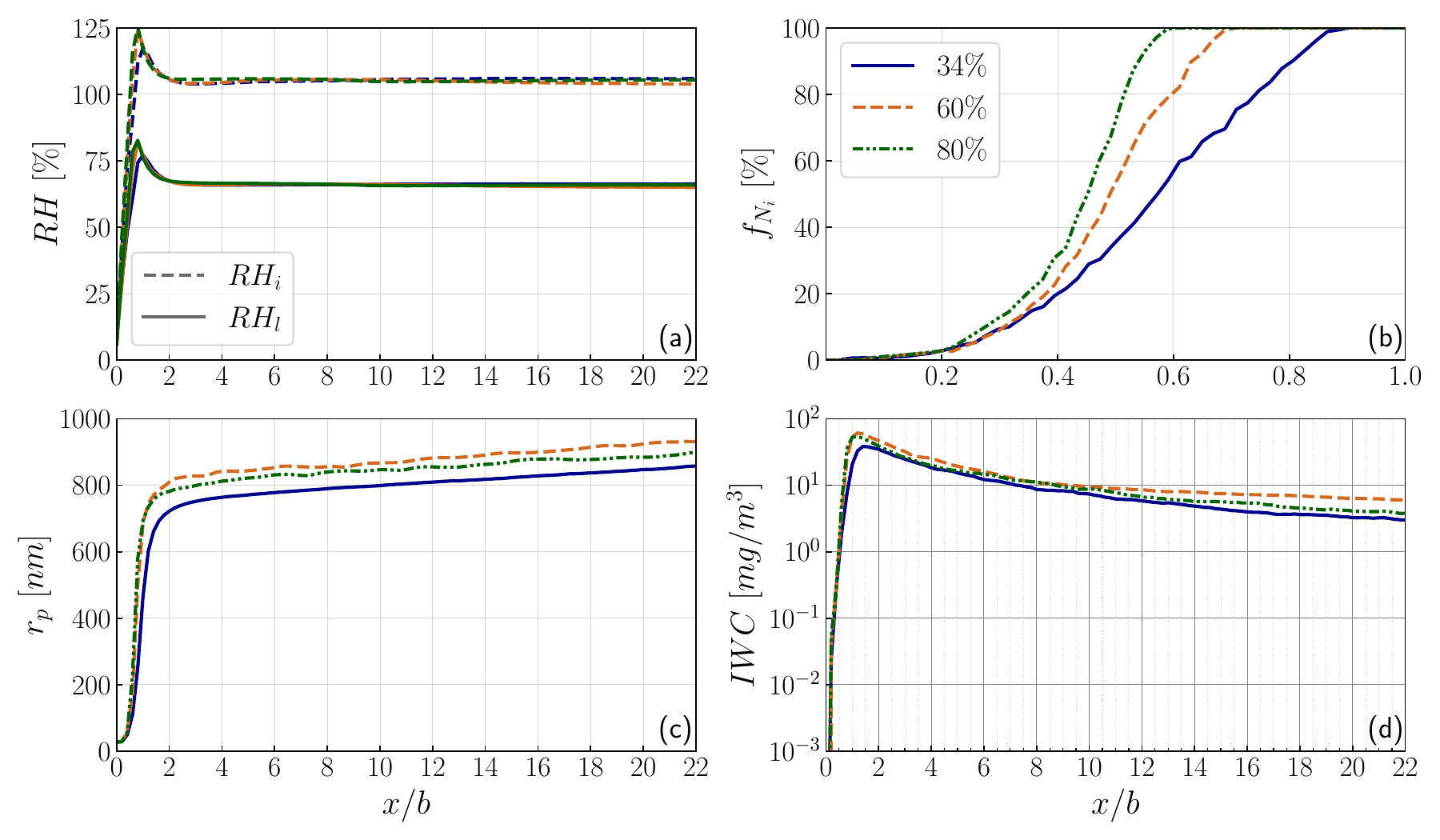}%
\end{center}
\vspace{-0.6cm}
\caption{Evolution of (a) relative humidity with respect to liquid water ($RH_l$, dashed lines) and to ice ($RH_i$, solid lines), (b) total ice crystal fraction $f_{N_i}$, (c) mean ice crystal radius $r_p$, and (d) mean ice water content $IWC$ as a function of the distance behind the aircraft.  In panel (b), the $x$-axis is limited to the interval 0 to 1 wingspan}
\label{fig:figure9}
\end{figure}

Modifying the engine position alters the dilution characteristics of the exhaust plumes in the aircraft wake. This, in turn, influences plume development on multiple levels, particularly affecting temperature and relative humidities conditions. These effects can be assessed by analyzing key quantities (\Cref{fig:figure9}), including relative humidities with respect to liquid water and ice (panel (a)), total ice crystal fraction (panel (b)), mean particle radius (panel (c)), and mean ice water content (panel (d)), for each engine configuration for the fully activated case. This scenario is used to analyze these effects because it enables an isolated analysis of water vapor diffusion within the plume and its dilution through mixing with ambient air in the absence of sulfur species. The relative humidities with respect to liquid water ($RH_l$) and ice ($RH_i$), the mean particle radius $r_p$ and the total ice crystal fraction $f_{N_i}$ are respectively defined as follows:

\begin{equation}
    RH_i = \frac{p_v}{p_{sat}^{ice}(\overline{T})}; \quad RH_l = \frac{p_v}{p_{sat}^{liq}(\overline{T})}
    \label{eq:equation18}
\end{equation}

\begin{equation}
    r_p = \frac{\sum_{k}r_{p,k}\,N_{p,k}\,V_k}{\sum_{k}N_{p,k}\,V_k}
    \label{eq:equation19}
\end{equation}

\begin{equation}
    f_{N_i}=\frac{\bar N_i }{\bar N_p}, \quad \bar N_i=\sum_k \mathds{1}_{r_{p}>1.01\times r_s} \,N_{p,k}\,V_k
    \label{eq:equation20}
\end{equation}

\noindent where $V_k$ is the volume of cell $k$, $\bar N_i$ is the number of ice crystals, $\mathds{1}_{r_{p}>1.01\times r_s}$ is the indicator function (equal to 1 when $r_{p}>1.01\times r_s$, and 0 otherwise), and $\bar N_p$ denotes the total number of soot particles, defined analogously to $\bar N_i$ but without applying the indicator function. All averages are computed over cells where $N_p>1\,m^{-3}$.

The pronounced differences between engine positions are primarily driven by dilution effects (\Cref{fig:figure4}). During the initial phase after ejection and jet expansion (up to one wingspan), the plumes in the 60\% and 80\% configurations undergo more rapid entrainment by the vortex. This accelerates the temperature drop, resulting in a higher relative humidities with respect to both liquid water and ice (panel (a)). The plumes reaching higher saturation ratio levels with respect to liquid water facilitate a faster onset of condensation, since in this model, condensation requires supersaturation with respect to liquid water to initiate (\Cref{eq:equation13}). It is noteworthy that the mean liquid water relative humidity ($RH_l$) remains near 80\% without exceeding the threshold in either case. This behavior is primarily attributable to the rapid expansion of the jet and the prompt onset of condensation, which quickly lowers the mean saturation ratio with respect to liquid water. Nevertheless, all the particles reach saturation locally, allowing for their activation across the plume.

Changes in relative humidities within the plume, driven by dilution, are subsequently reflected in condensation behavior. Consequently, ice crystals form and grow more rapidly in configurations closer to the wingtip vortex, namely 60\% and 80\%, compared to the 34\% case, closer to the fuselage (panels (b) and (c)). Further downstream, the differences in crystal radii remain consistent, despite the change in dilution trends (\Cref{fig:figure4}). The supersaturation differences established within the first wingspan persist, as the plume remains saturated with respect to ice in all three cases ($RH_i > 100\%$), continuously supplying water vapor to support ice crystal growth. Finally, the 60\% configuration produces the largest crystals, followed by the 80\% and 34\% configurations. This trend is consistent with the findings of Ramsay et al. \cite{RR2024}, although their results indicate slower crystal growth as compared to the present study. Comparisons with other studies reporting ice crystal radii for realistic configurations show good agreement with the magnitudes obtained herein, particularly near the engine, as reported by Cantin et al. \cite{CANTIN22}. The findings of Khou et al. \cite{KHOU2016} are also relevant. Their highest sulfur content case, which results in a fully activated surface fraction ($\theta_a$ = 100\%), yields crystal sizes similar to those in the fully activated scenario. A similar trend in ice crystal radius magnitudes is reported in the study by Bier et al. \cite{BIER22}, which employed a box model incorporating soot and activation based on $\kappa$-Köhler theory. Differences between our cases and the literature can be attributed to variations in modeling, engine, and atmospheric conditions.

These variations in saturation ratio and ice crystal radius ultimately influence the ice water content of each contrail (panel (d)), which follows similar trends to those observed for the mean ice crystal radius across the different engine configurations. Interestingly, the 80\% configuration does not yield the highest radii and ice water content. This outcome is primarily attributed to enhanced dilution-driven diffusion of water vapor within the plume, which leads to a more rapid decrease in relative humidity compared to the 60\% configuration.

\subsubsection{Comparison between engine position and relative humidity sensitivities}

\begin{figure}[ht]
\begin{center}
\includegraphics[width=1.\textwidth]{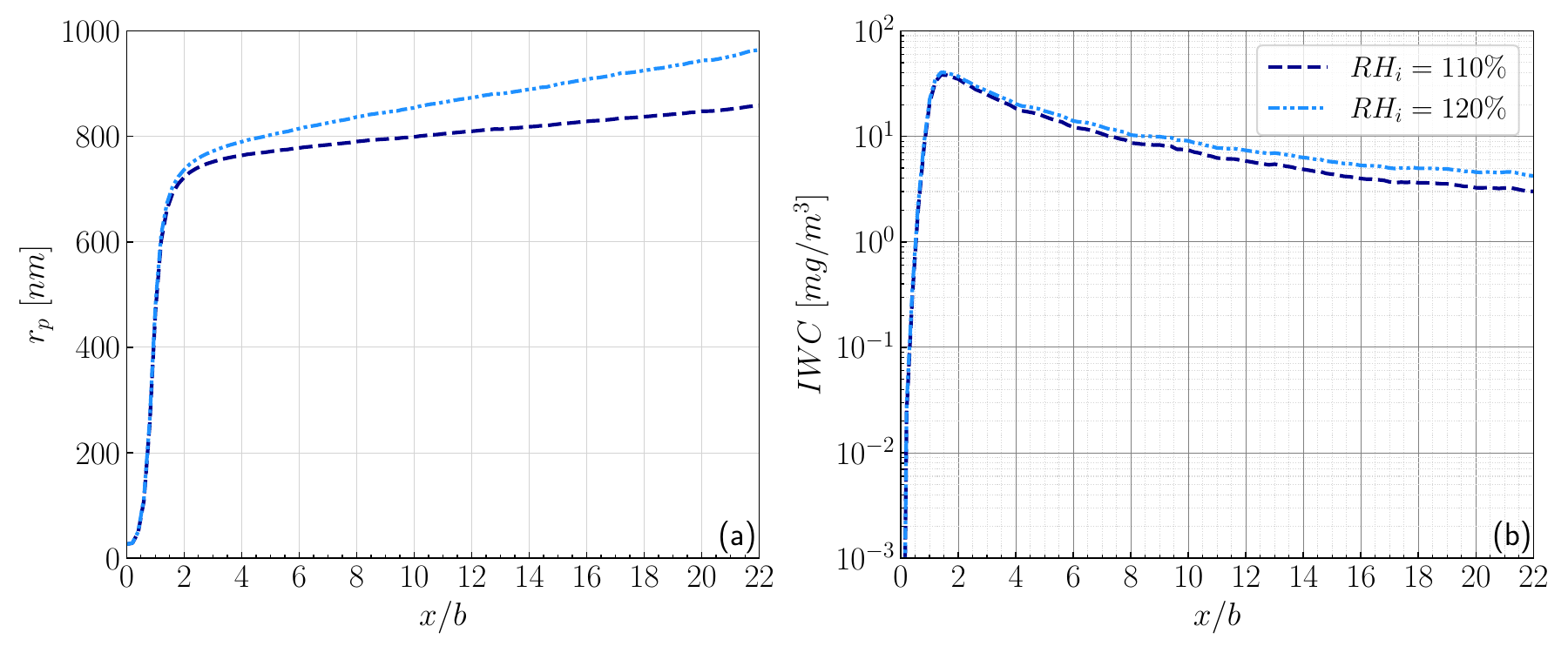}%
\end{center}
\vspace{-0.6cm}
\caption{Evolution of (a) mean ice crystal radius $r_p$, and (b) mean ice water content $IWC$ as a function of the distance behind the aircraft, for ambient $RH_i=110\%$ (dashed lines) and $120\%$ (dashed-dotted lines)}
\label{fig:figure10}
\end{figure}

Ambient relative humidity and temperature are primary parameters governing the formation and evolution of contrails \cite{Kar09}. In particular, relative humidity plays a critical role by continuously supplying water vapor to the developing contrail throughout its lifespan. It is therefore insightful to assess the sensitivity of contrail formation to variations in relative humidity, especially in comparison to the sensitivity associated with engine position, as seen in the previous section. This sensitivity is investigated by two scenarios: one with a relative humidity of 110\%, as previously analyzed, and another with a more supersaturated environment at 120\% for the 34\% case. The comparison is presented in Figure \ref{fig:figure10}, which shows the evolution of both the ice crystal radius (Equation \ref{eq:equation20}) and the $IWC$ under these two ambient humidity values.

As expected, the results clearly indicate that an increase in ambient relative humidity leads to a corresponding rise in both the mean ice crystal radius and the contrail’s ice water content. Notably, for the slope of the ice radius curve, enhanced availability of water vapor promotes more rapid depositional growth, thereby increasing the total ice mass within the contrail. This shift in the growth dynamics is of particular interest here, as it provides a basis for comparison with the sensitivity to engine position. Prior to a distance of two wingspans downstream of the aircraft, the values remain essentially identical across ambient humidity scenarios, with water vapor from the engine jet plume primarily governing ice crystal growth. The differences in mean particle radius between the 110\% and 120\% relative ambient humidity cases are then approximately 2\% for $r_p$ and 6\% for the $IWC$. In contrast, variations arising from changes in engine position are substantially larger: for the 60\% configuration, differences reach about 12\% in $r_p$ and 35\% in $IWC$, while the 80\% configuration yields differences of roughly 8\% and 12\%, respectively. Further downstream, at a distance of 20 wingspans, the impact of increased humidity becomes more pronounced. The 120\% relative humidity case exhibits an increase of approximately 11\% in $r_p$ and 40\% in $IWC$ relative to the 110\% baseline. The sensitivity to engine position remains strong: the 60\% configuration shows differences of around 9\% in $r_p$ and 95\% in $IWC$, whereas the 80\% case shows respective increases of 4.5\% and 26\%, both compared to the reference 34\% configuration.

Given these differences, modifying the engine position leads to a more rapid initial growth of ice crystals in the near-field region than would be induced by a 10\% change in ambient relative humidity. This effect arises primarily from enhanced plume mixing in the modified configurations, which accelerates cooling and the amount of available water vapor for condensation. Further downstream, however, ambient relative humidity becomes the dominant factor, as it governs the sustained supply of water vapor to the contrail, progressively amplifying differences in crystal growth. While changes in engine position produce effects on ice particle radius that are of comparable magnitude to those induced by relative humidity variations, the resulting changes in ice water content can vary more strongly, depending on the specific engine configuration, though they remain broadly of the same order. Overall, in the near field, the impact of engine position changes is roughly equivalent to that of a 10\% variation in ambient relative humidity, underscoring the role of dilution in shaping the early stages of contrail evolution. In light of the observed sensitivities, variations in ambient temperature would likely exert an even more pronounced influence on contrail microphysics.

\subsubsection{Sensitivity to the soot number emission index}

\begin{figure}[ht]
\begin{center}
\includegraphics[width=1.\textwidth]{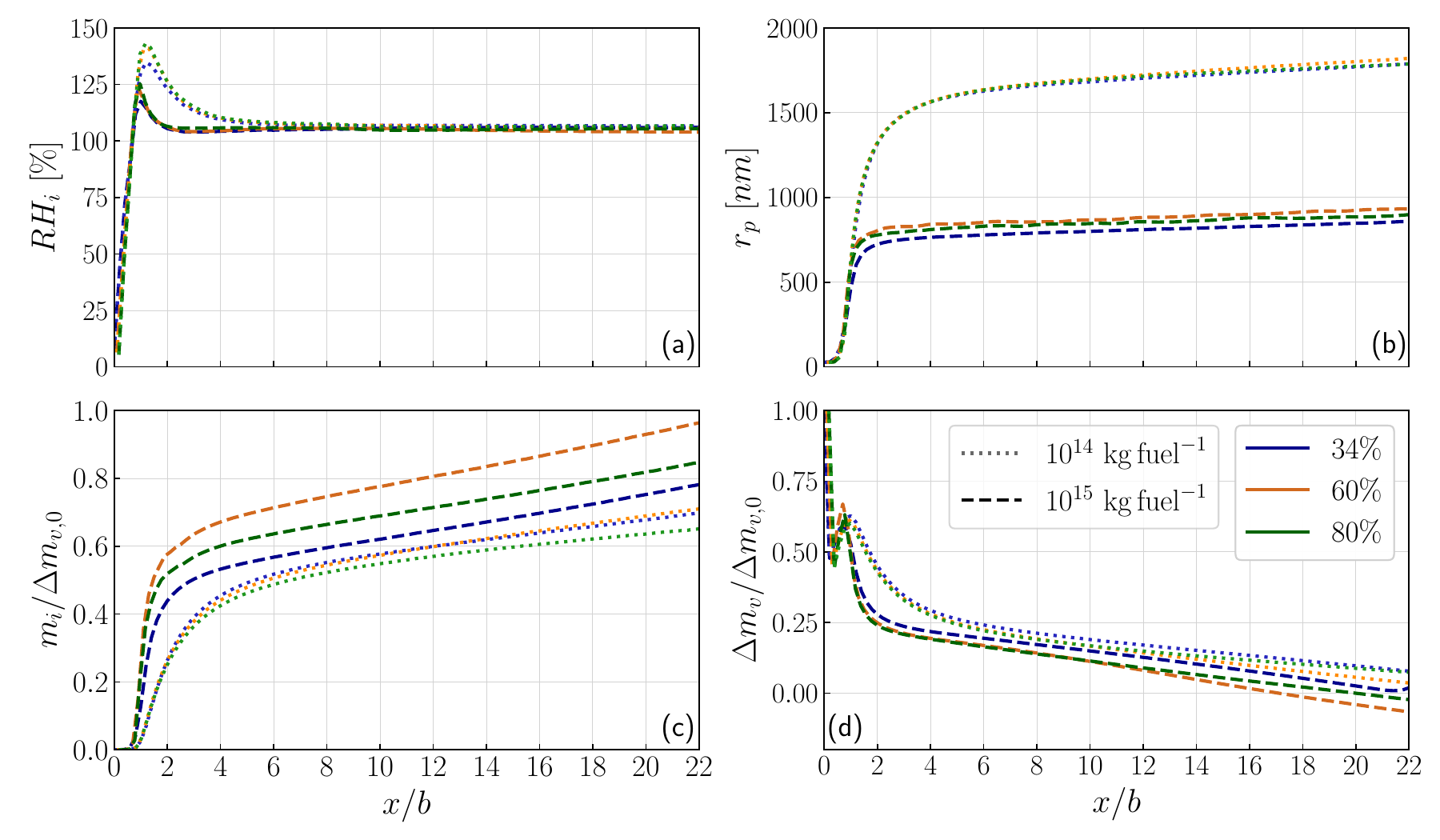}%
\end{center}
\vspace{-0.6cm}
\caption{Evolution of (a) relative humidity with respect to ice $RH_i$, (b) mean ice crystal radius $r_p$, (c) ice mass $m_i$ and (d) excess of water vapor $\Delta m_v$ as a function of the distance behind the aircraft, for $\text{EI}_{\text{soot}}=10^{14} \text{ kg-fuel}^{-1}$ (lighter dotted lines) and $\text{EI}_{\text{soot}}=10^{15} \text{ kg-fuel}^{-1}$ (darker dashed lines)}
\label{fig:figure11}
\end{figure}

The soot number emission index at the engine outlet is a key parameter in contrail formation \cite{Kar09}. It directly governs the number of available nucleation sites for ice crystals \cite{KLEI18} and significantly influences condensation dynamics by leading to competition for water vapor. A higher soot particle concentration enhances this competition, whereas a lower concentration allows more vapor to condense per particle. As highlighted in Paoli et al. \cite{PAO2013}, condensation in the jet regime is driven by two key mechanisms: turbulent mixing and vapor depletion. The first process is particularly sensitive to engine position, which, as shown in the previous section, strongly influences the mixing characteristics of the plume. The sensitivity of the second mechanism to engine position is then investigated by comparing two soot number emission indices, $\text{EI}_{\text{soot}}=10^{14} \text{ kg-fuel}^{-1}$ and $\text{EI}_{\text{soot}}=10^{15} \text{ kg-fuel}^{-1}$, under the fully activated scenario. The results are presented in \Cref{fig:figure11}. Panel (a) shows the relative humidity with respect to ice $RH_i$ (\Cref{eq:equation18}), panel (b) the mean ice crystal radius $r_p$ (\Cref{eq:equation19}), panel (c) the ice mass $m_i$ and panel (d) the excess of water vapor in the plume $\Delta m_v$. Both the ice mass and the vapor excess are normalized by the total water vapor released by the engine $\Delta m_{v,0}$. These quantities are defined following Paoli et al. \cite{PAO2013}:

\begin{equation}
    m_i(x)=\iint \rho Y_{H_2O,s}dS
    \label{eq:equation21}
\end{equation}

\begin{equation}
    \Delta m_v(x)=\iint \rho (Y_{H_2O}-Y_{H_2O,\infty})dS
    \label{eq:equation22}
\end{equation}

\noindent where $Y_{H_2O,\infty}$ denotes the ambient mass fraction of water vapor. All quantities are computed over grid cells where the particle number density satisfies $N_p>1\,m^{-3}$.

Across the two soot number emission indices, similar trends are observed with respect to plume ice saturation ratio (panel (a)) and mean ice crystal radius (panel (b)) across engine positions. As noted in the previous section, the initial wingspans downstream of the engine exhibit consistent mixing behavior regardless of the emission index. Specifically, configurations closer to the wingtip vortex experience more rapid cooling and achieve higher levels of saturation ratio. This results in both earlier contrail formation and the development of larger ice crystals in the 60\% and 80\% engine placement cases. Naturally, the absolute levels differ between the two emission indices: with fewer soot particles, more water vapor is available for condensation per particle, reducing competition and leading to the formation of larger ice crystals.

However, it is beyond the very near field that differences in trends between emission indices become apparent. At the lower emission index, the curves converge more closely, exhibiting smaller variations across engine positions compared to the higher emission index case. This effect can be understood by examining the evolution of ice mass and excess water vapor in the plume (panels (c) and (d)). As shown by Paoli et al. \cite{PAO2013}, a higher concentration of soot particles leads to more rapid depletion of water vapor, along with a greater rate of conversion to ice. Concerning engine position, configurations closer to the wingtip experience more intense mixing, which further accelerates vapor depletion. This depletion effect is therefore amplified under high soot number concentrations, where mixing rapidly controls the plume’s water supply. In contrast, at lower soot number concentrations, the available water vapor, primarily from engine exhaust, sustains ice crystal growth for a longer period. Crystals thus benefit from more favorable growth conditions and deplete the available water more gradually. As a result, dilution-related differences between engine positions become less pronounced at low emission indices than under high particle loading. These findings suggest that soot number concentration plays a more critical role than engine placement in governing contrail formation.

\subsubsection{Impact on soot activation}

\begin{figure}[ht]
\begin{center}
\includegraphics[width=1.\textwidth]{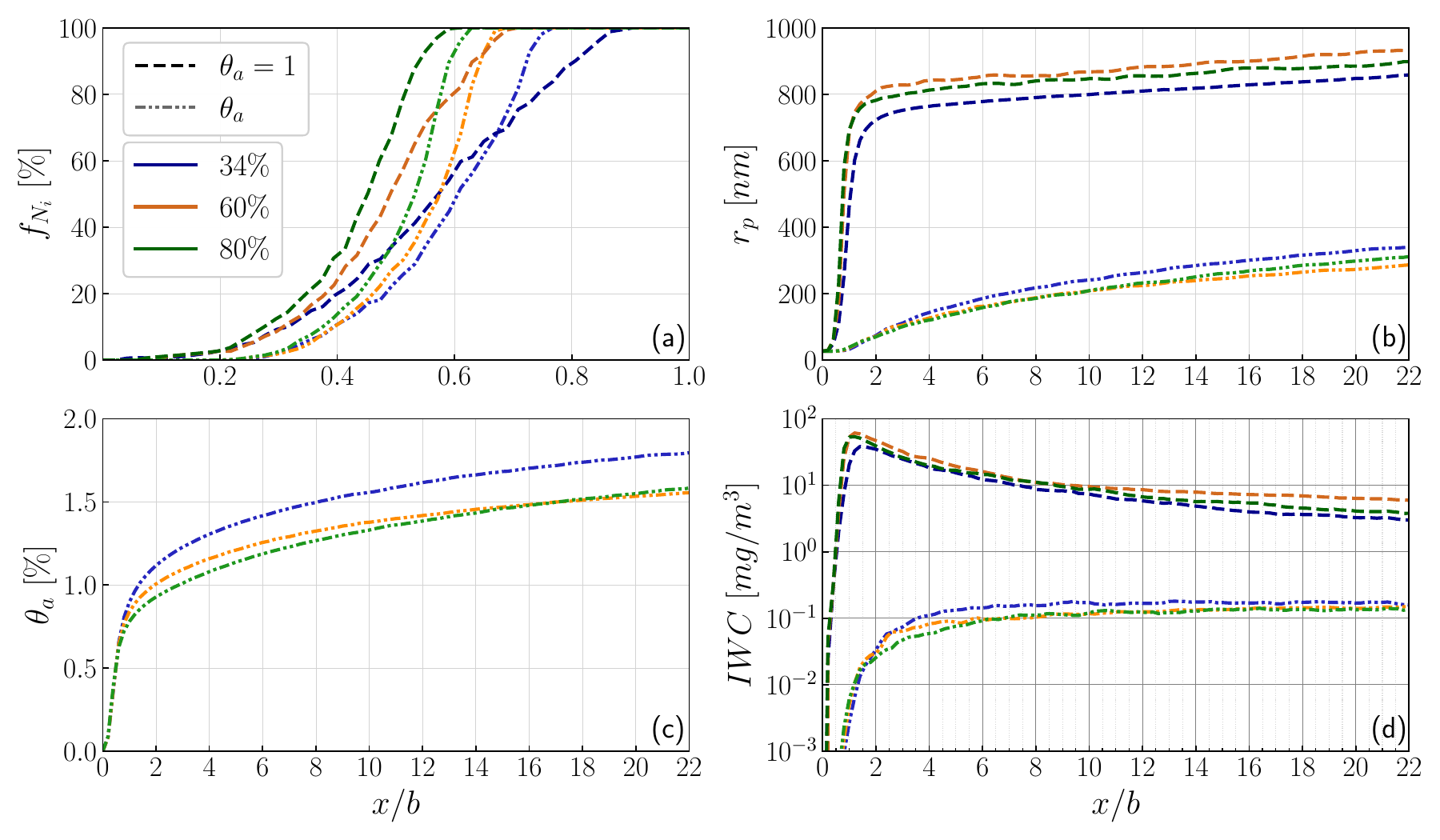}%
\end{center}
\vspace{-0.6cm}
\caption{Evolution of (a) total ice crystal fraction $f_{N_i}$, (b) mean ice crystal radius $r_p$, (c) mean surface fraction of activation $\theta_a$, and (d) mean ice water content $IWC$ as a function of the distance behind the aircraft for the scenarios fully activated (darker dashed lines) and with activation through adsorption (lighter dash-dotted lines). In panel (a), the x-axis is limited to the interval 0 to 1 wingspan}
\label{fig:figure12}
\end{figure}

The engine position and the associated variation in plume dilution not only affect relative humidity levels but also modulate the dilution of chemical species within the aircraft wake. This is particularly relevant for compounds critical to soot particle activation, such as H\textsubscript{2}SO\textsubscript{4} and SO\textsubscript{3}. Accelerated entrainment of the plume leads to a more expanding plume (\Cref{fig:figure8}) and rapid mixing with ambient air. This leads to a stronger dilution of activating species, thereby directly reducing particle activation efficiency. This effect is illustrated in \Cref{fig:figure12}, where the fully activated scenario (denoted $\theta_a=1$) is compared with the scenario involving sulfur-driven activation (denoted $\theta_a$). The total ice crystal fraction $f_{N_i}$ (\Cref{eq:equation20})  is shown in panel (a), the mean ice crystal radius $r_p$ (\Cref{eq:equation19}) in panel (b), the activated soot surface fraction (for the sulfur-activation case) $\theta_a$ in panel (c), and the mean ice water content $IWC$ in panel (d), for each engine position and activation scenario.

The presence of sulfur species does not significantly alter the overall onset of condensation behavior across the three engine configurations (panel (a)). In both scenarios, ice crystal nucleation occurs most rapidly in the 80\% configuration, followed by the 60\% and then the 34\% cases. This trend is primarily driven by faster saturation in configurations closer to the wingtip vortex, as seen in \Cref{fig:figure9}. However, despite this favorable saturation, ice crystal nucleation proceeds more slowly in the presence of sulfur species, due to the limited surface fraction of activated soot particles (\Cref{eq:equation10}), which restricts condensation to the activated surfaces. However, the early-stage dilution, occurring within the first wingspan, on sulfur species influences the evolution of the other microphysical properties in the scenario with adsorption activation. For this scenario, the 34\% configuration produces the largest crystals, followed by the 80\% and 60\% configurations, reversed compared to fully activated cases (panel (b)). The initial dilution then reduces the concentration of sulfur species, leading to a lower activation fraction, as shown in panel (c). As a result, the differences in sulfur species dilution yield smaller crystals in the 60\% and 80\% configurations than 34\%. This inversion is primarily driven by the enhanced diffusion of sulfur species, which becomes the dominant factor limiting particle activation and subsequent ice crystal growth. These trends underscore both the role of engine placement in species dilution and the sensitivity to microphysical modeling assumptions. These radii magnitudes align with the low-sulfur case reported by Khou et al. \cite{KHOU2016}, which examines the effects of varying sulfur concentrations. The surface fraction of activation aligns with the orders of magnitude for low FSC in Wong et al. \cite{WONG10}. However, the ice crystals remain relatively small, as ice condensation is constrained by the limited fraction of soot particle surface area that becomes activated (\Cref{eq:equation11}). This effect is primarily driven by the low sulfur content at the engine outlet and the absence of a term accounting for the effect of condensation on particle activation, as incorporated in the work of Wong et al. \cite{Wong14}. Finally, these variations impact the ice water content of each contrail (panel (d)), where, as in the fully activated case, the trends in crystal radius across engine configurations remain consistent. However, under sulfur-driven activation, the ice content levels are notably closer between configurations. This indicates that activation effects, and, by extension, the microphysical modeling, seems to have a greater impact on contrail formation than engine position in the context of this study.

\subsubsection{Effect of contrail structure on ice crystal growth}

\begin{figure}[ht!]
\begin{center}
\includegraphics[width=1.\textwidth]{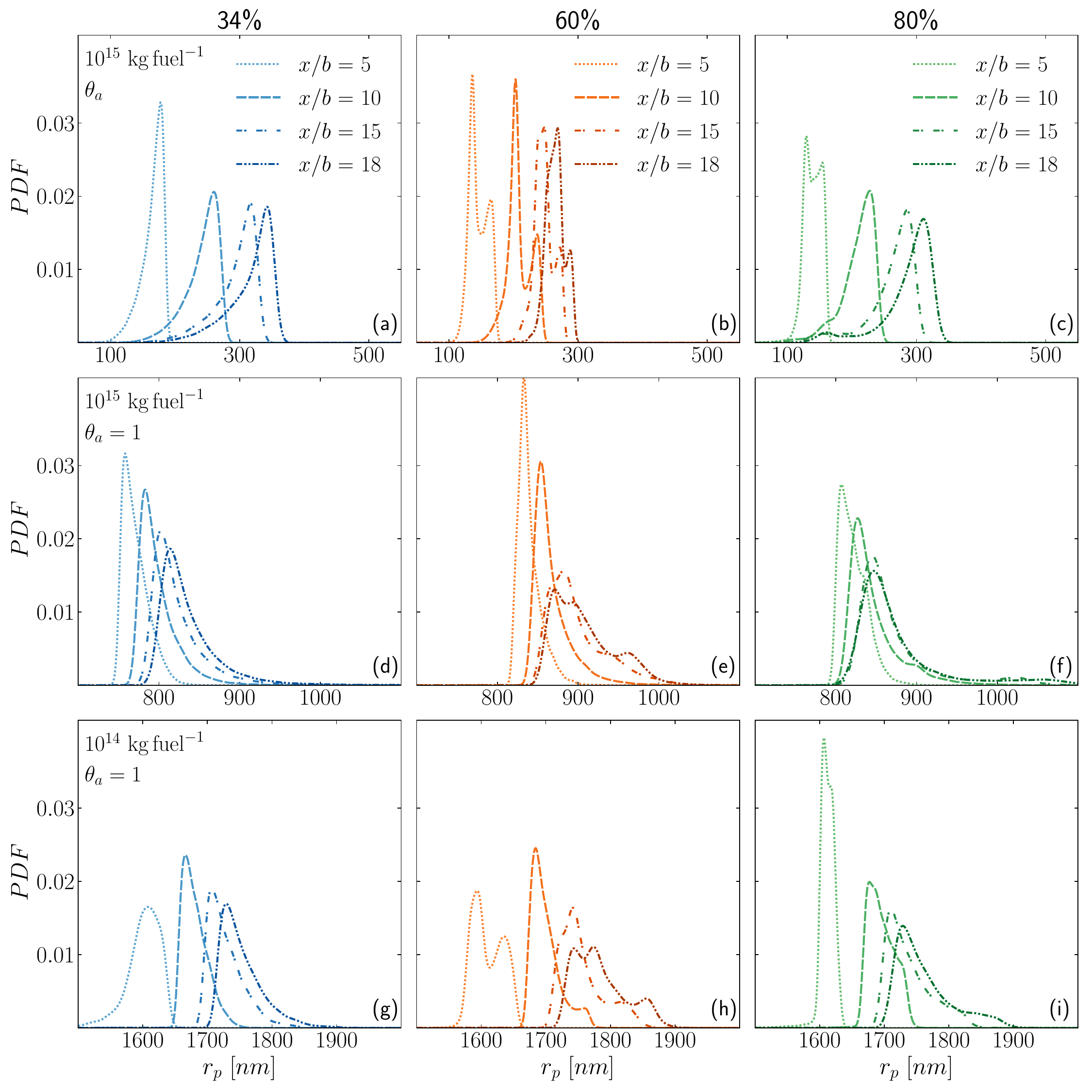}%
\end{center}
\caption{Evolution of particle radius PDFs for the configurations (34\%, 60\%, 80\%): (a-c) activation through adsorption, (d-f) fully activated with $\text{EI}_{\text{soot}}=10^{15} \text{ kg-fuel}^{-1}$, and (g-i) fully activated with $\text{EI}_{\text{soot}}=10^{14} \text{ kg-fuel}^{-1}$ at distances $x/b$ = 5, 10, 15, and 18 downstream. The x-axis is scaled according to the characteristic radius associated with each activation scenario}
\label{fig:figure13}
\end{figure}

The structure of each contrail, influenced by engine displacement and directly linked to aerodynamic factors, plays a crucial role in the growth of ice crystals. Variations in the mean particle radius are further emphasized when analyzing the probability density functions (PDFs) of radius for each configuration, both with activation through adsorption and fully activated, as well as both soot number emission indices. (\Cref{fig:figure13}). The left column corresponds to the 34\% configuration, the center to 60\%, and the right to 80\%. These plots reveal the evolution of crystal sizes during contrail formation, offering a clearer understanding of the spatial distribution differences between the configurations.

For each configuration and scenario, the progressive growth of ice crystals can be observed as the distance behind the aircraft increases. In terms of magnitude, the crystal radius distributions vary significantly between scenarios. As discussed in previous sections, the differences between the two soot number emission indices arise from variations in the number of available nucleation sites and the associated water vapor competition. In contrast, under adsorption-based activation, ice crystal growth remains limited due to the reduced surface activation fraction (\Cref{fig:figure12}), which constrains subsequent condensation. Nonetheless, the observed trends are insightful, as they emphasize the role of engine placement in shaping the dilution dynamics of both sulfur species and water vapor, factors that critically influence ice crystal growth within the evolving plume.

For the scenario with activation through adsorption (panels (a-c)), the growth trends are similar for the 34\% and 80\% cases, where a prominent peak in crystal radius, of the same order of magnitude, is observed at each wingspan. This behavior aligns with the larger crystal sizes observed in the 34\% case. In contrast, the 60\% case exhibits two distinct peaks, arising from the portion of the contrail that penetrates the vortex. This penetration creates a unique contrail distribution, where some ice crystals grow independently of those in the jet, driven by water vapor entrained into the vortex. A small plateau at smaller radii is also visible in the 80\% case, similar to the 60\% case, due to the jet's partial presence within the vortex. However, in the 80\% case, the extent of jet penetration into the vortex is smaller, leading to a different growth trend as compared to the 60\% case. 

For the fully activated scenarios (panels (d-i)), the trends are similar among themselves but differ slightly from those observed in the scenario involving activation through sulfur species adsorption. Crystal radii are already larger at earlier stages, depleting water vapor more quickly and leaving less available for further growth. This effect is particularly evident when examining the relative humidity with respect to ice, which is lower in this scenario (\Cref{fig:figure9}) than in the case of adsorption activation. Consequently, areas with the largest radii correspond to regions with fewer particles: those penetrating the vortex for the 60\% case (second peak close to 1000 nm) and those entrained into the vortex dipole for the 80\% case (small plateau around 1000 nm). For the 34\% configuration, particles are concentrated in the jet as it wraps around the vortex, leaving those at the jet’s edge to grow the most, as water vapor is rapidly depleted within the core. 

These observations are particularly significant as they emphasize the impact of the contrail’s spatial distribution on ice crystal growth. Depending on their location—whether within the jet, the vortex, or its surroundings—and the activation scenario, ice crystals exhibit distinct growth patterns due to variations in water vapor supply. These results, along with those from the previous two sections, complement the findings of Paoli et al. \cite{PAO2013} and Ramsay et al. \cite{RR2024}, enabling a more in-depth analysis that highlights the significant impact of engine position on contrail formation and microphysical processes. Depending on the engine position, particles experience different thermodynamic conditions, which will subsequently affect their evolution and radiative impact.

\subsection{Impact on contrail radiative properties}

\begin{figure}[H]
\begin{center}
\includegraphics[width=1.\textwidth]{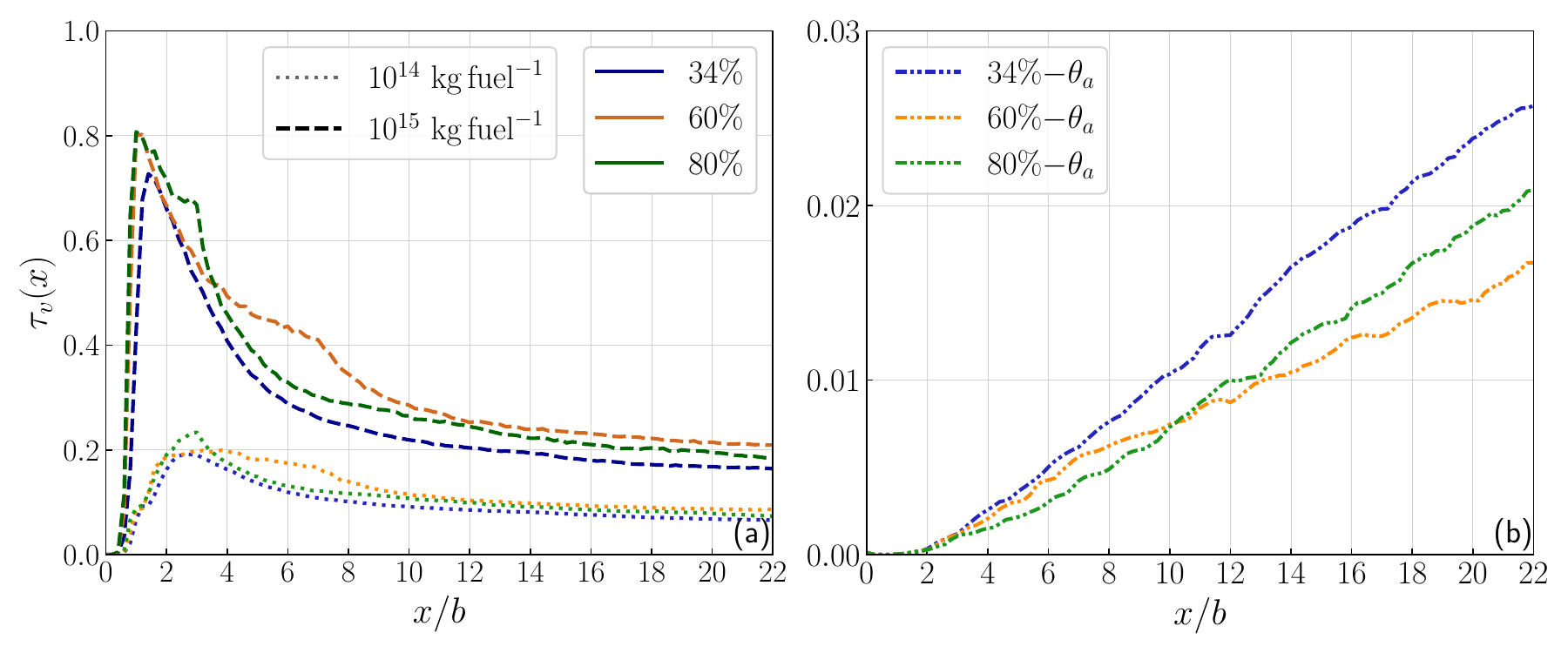}%
\end{center}
\vspace{-0.6cm}
    \caption{Evolution of the mean optical thickness $\tau_v(x)$ for the three engine configurations as a function of distance behind the aircraft: (a) fully activated scenarios with $\text{EI}_{\text{soot}}=10^{14} \text{ kg-fuel}^{-1}$ (lighter dashed lines), and with $\text{EI}_{\text{soot}}=10^{15} \text{ kg-fuel}^{-1}$ (darker dashed lines), and (b) activation through adsorption (lighter dash-dotted lines)}
     \label{fig:figure14}
\end{figure}

The mean optical thickness of the plume, $\tau_v(x)$ (\Cref{eq:equation23}), is plotted along the aircraft's trajectory (\Cref{fig:figure14}). This value quantifies the environment's transparency, with higher optical thickness indicating a denser contrail. It is then directly related to the contrail's radiative impact. It is computed by interpolating the RANS solutions onto a Cartesian grid for each configuration, enabling the required integration to be performed. The Mie coefficient $Q_{ext}(r_p)$, included in the $\tau_v(x)$ equation, is defined in \cref{eq:equation24} \cite{qext}, where $\lambda_w = 0.55~\mu m$ and the refractive index of ice is $\mu_r=1.31$.

\begin{equation}
\tau_v(x) = \int_{-\infty}^{+\infty} \pi r_p^2 N_p Q_{ext}(r_p)dz
\label{eq:equation23}
\end{equation}

\begin{equation}
Q_{ext}(r_p) = 2 - \frac{4}{e}\bigg(sin(e) - \frac{1-cos(e)}{e}\bigg); \quad e = \frac{4\pi r_p(\mu_r-1)}{\lambda_w}
\label{eq:equation24}
\end{equation}

The optical thickness for the three cases in each activation scenario shows significant variation. In the fully-activated cases, the 60\% configuration has the highest value, followed by the 80\% and 34\% configurations for soot number emission index. Conversely, in the activation cases, the order is reversed, the 34\% configuration exhibits the highest value at the end of the domain, followed by the 80\% and 60\% configurations. This behavior is primarily driven by the evolution of ice crystal size, which follows a similar trend (\Cref{fig:figure11,fig:figure12}). Larger and more numerous ice crystals contribute to an increase in optical thickness within the contrail, as shown by \cite{Kar09}. A higher optical thickness can also be attributed to an increased soot density caused by reduced dilution (\Cref{fig:figure4}). The combination of the two then gives these trends. It is interesting to note that the mean optical thickness values obtained in this study align well, in terms of magnitude, with those reported by Khou et al. \cite{KHOU2016} and follow the trends observed, for engine position, by Ramsay et al. \cite{RR2024} in fully activated scenarios. In the case of activation by sulfur species, the resulting optical thickness remains significantly lower than values typically reported in the literature \cite{Schu17}. As previously discussed, this outcome is primarily driven by the low fraction of soot surface activated, which constrains ice condensation and ultimately results in the formation of smaller ice crystals.

Regarding magnitude, greater confidence should be placed in the fully activated scenarios. Nevertheless, this scenario remains valuable for analyzing broader trends. It highlights the full extent of dilution effects on soot-activating species, in contrast to the fully activated cases, where dilution impacts mainly water vapor. This allows for a more nuanced understanding of how engine position and associated dilution patterns affect contrail radiative properties. For instance, under sulfur-driven activation, the 34\% configuration exhibits a higher optical thickness, whereas it shows the lowest value when sulfur activation is not considered, emphasizing the sensitivity of radiative effects to the activation pathway. These trends are significant as they provide insight into the impact and, more importantly, the state of contrails at the end of the jet regime. However, they do not determine the optimal engine position, as the later evolution of the contrail is just as crucial as its formation process. Nevertheless, they demonstrate the influence of engine position from the earliest stages of contrail development.

\section{Conclusion}
\label{sec:section6}

\subsection{General synthesis}

Simulations of contrail formation and evolution up to 22 wingspans downstream of the aircraft were performed using a RANS and Eulerian bulk approach on a complete and realistic 3D airliner configuration. These simulations examined three distinct engine positions, considering both soot activation through adsorption and fully activated scenarios, the latter evaluated for two different soot number emission indices. By employing a mesh adaptation technique, the vortex sheet, wake vortex, and turbulent jet were accurately captured, enabling a detailed analysis of jet-vortex interactions and their effects on contrail formation across all configurations. This study provides a more detailed exploration of the interaction between aerodynamic and microphysical phenomena, building upon the work of Ramsay et al. \cite{RR2024}. The results revealed significant aerodynamic changes induced by engine displacement. In the 60\% and 80\% configurations, new vortex structures emerged, which were not present in the 34\% configuration, where the engine is positioned closer to the fuselage. These new structures appeared to arise from the differentiation of the vortex sheet, influenced by the engine displacement, which led to flow separation on the wing.

Altering the engine position also leads to variations in jet dilution: 34\% for the most diluted case in terms of particle density, followed by 80\% and 60\%, near the end of the computational domain. It directly influences the characteristics of the contrails formed in each configuration. The cross-sectional area of the contrail for the three configurations follows a power law, with values consistent with the orders of magnitude reported in the literature \cite{SCHU98}. The scaling coefficient, however, is dependent on the engine configuration. The spatial distribution of particles, but also the overall structure of the contrails, changes.  As a result, distinct properties emerge, including differences in ice crystal size and relative humidities levels within the plume. 

The effects of dilution on microphysical properties are evident at multiple levels. Firstly, on plume saturation ratio, the earlier entrainment of the jet induces temperature differences, leading to higher relative humidities levels within the plume and promoting faster ice crystal growth. Consequently, these differences persist throughout the contrail’s evolution, as it continues to be sustained by high relative humidity (RH = 110\%). In the presence of a lower number of soot particles, this effect remains noticeable during the early stages but does not persist further downstream, primarily due to competition for water vapor, which becomes depleted rather than diluted. Soot number concentration appears to have a stronger influence on contrail formation than engine position. In the case of activation through adsorption, dilution also affects the concentration of species responsible for soot surface activation. As a result, configurations with the highest dilution have the lowest surface activation fractions. This has the effect of reversing the previously observed trends, where the largest particle sizes are no longer observed for the same configurations. The differences in engine position effects across all scenarios directly influence the optical properties of the contrail, particularly the mean optical thickness. 

Based on the simulation results of this study, engine position appears to influence the properties of the resulting contrails. This effect is, at least in the near-field, of comparable magnitude to that resulting from a 10\% increase in ambient relative humidity. However, this effect seems to be relatively less significant compared to first-order parameters such as the soot number emission index. The results also show a notable sensitivity to the activation mechanism employed. Finally, it further highlights the importance of accurately resolving the mixing processes. Employing a realistic 3D configuration is particularly relevant for this purpose, as it captures the full dynamics behind the aircraft despite the associated computational costs.

\subsection{Discussion and perspectives}

This study on the influence of engine position on contrail formation provides valuable insight into overall trends and dilution mechanisms. Nevertheless, it is important to emphasize that the proposed engine relocation is a purely theoretical and academic consideration. In practice, such modifications would likely entail significant structural constraints and increased weight penalties, consequently altering the overall aerodynamic circulation of the aircraft. Considering the assumptions involved in this analysis, the findings provide useful insights but should be interpreted within the framework of the study’s assumptions.

One of the primary limitations of this study is the occurrence of flow separation induced by engine positioning. This limitation arises from the use of a realistic aircraft configuration with a wing that is not optimized for engine repositioning. While the wake vortex remains the dominant factor in jet-vortex interaction, flow separation introduces an unintended effect when accounting for aircraft geometry. Investigating a scenario where the wing is optimized for each engine position could offer a valuable direction for future research.

Furthermore, the choice of RANS simulations to account for aircraft geometry is significant but comes with inherent limitations. The unsteadiness of the flow, including likely instabilities between the highly turbulent jet and the forming vortex, is lost. Small-scale flow structures are not explicitly resolved but are instead modeled through a turbulence model. This has implications for contrail dynamics, which should be considered. For example, using a $k$-$\omega$ model results in a larger vortex radius compared to a DRSM model \cite{CHURCH09}, potentially leading to slight differences in contrail dispersion.

Additionally, microphysical assumptions play a significant role in accurately simulating contrail formation. The first key point concerns the influence of the particle activation mechanism employed in this study, particularly when compared to scenarios without activation. Under low FSC conditions, crystal growth is significantly constrained relative to the fully activated case, resulting in smaller particle radii and markedly reduced optical thickness. This highlights the need to account for condensation effects on activation, as incorporated in the work of Wong et al. \cite{Wong14}. Therefore, we conclude that the fully activated case is probably the most relevant in this study. For future work, it would be valuable to incorporate the activation due to condensation into the microphysical model to enable a more robust assessment of engine position effects in scenarios where soot particles require activation to form ice crystals. Exploring dilution effects with more detailed microphysics or in polydisperse cases would then be valuable. Polydispersion may then alter the activation dynamics of particles by sulfuric compounds, as the governing parameters are dependent on soot radius (\Cref{eq:equation9,eq:equation10}). Consequently, smaller particles tend to be more activated than larger ones under a given sulfur concentration. This size distribution also influences ice crystal activation and condensation processes, wherein larger particles are thermodynamically favored for water uptake. As a result, competition for available water vapor intensifies, a relevant factor when assessing the impact of engine placement, as it directly affects local plume relative humidities levels. The use of more comprehensive particle activation pathways, as in Cantin et al. \cite{Cantin25}, could also be of interest, given the observed effect of dilution on particle activation in this study. Moreover, the sensitivity of engine position to atmospheric temperature is not addressed in the present study. This choice has been motivated by the objective of exploring aircraft design strategies aimed at mitigating contrail formation. Given that ambient temperature is a critical parameter governing contrail formation \cite{Kar09}, its interplay with engine placement warrants dedicated investigation.

The limitations discussed above also point to promising directions for future research. The inclusion of near-field effects and aircraft geometry, as demonstrated in this study, is crucial, as many phenomena occurring in this region significantly influence contrail formation. Then, regarding future work, it would also be valuable to investigate the longer-term evolution of contrails, particularly during the vortex phase, to determine whether the differences induced in the near field persist. This could be accomplished by employing methods that enable the transition from a RANS framework to a LES approach \cite{YOU24}, allowing to capture instability phenomena that disrupt the vortex pair. 

\section*{Acknowledgments}

The authors would like to sincerely thank Marc Massot for his valuable feedback and insightful suggestions, which have significantly contributed to improving the clarity and quality of this article. 

\bibliographystyle{elsarticle-num} 
\bibliography{cas-refs}

\begin{thebibliography}{10}
\expandafter\ifx\csname url\endcsname\relax
  \def\url#1{\texttt{#1}}\fi
\expandafter\ifx\csname urlprefix\endcsname\relax\def\urlprefix{URL }\fi
\expandafter\ifx\csname href\endcsname\relax
  \def\href#1#2{#2} \def\path#1{#1}\fi

\bibitem{LEE2009}
D.~S. Lee, D.~W. Fahey, P.~M. Forster, P.~J. Newton, R.~C. Wit, L.~L. Lim,
  B.~Owen, R.~Sausen, Aviation and global climate change in the 21st century,
  Atmospheric Environment 43~(22) (2009) 3520--3537.
\newblock \href {https://doi.org/10.1016/j.atmosenv.2009.04.024}
  {\path{doi:10.1016/j.atmosenv.2009.04.024}}.

\bibitem{BURK2011}
U.~Burkhardt, B.~Kärcher, Global radiative forcing from contrail cirrus,
  Nature Clim Change 1 (2011) 54--58.
\newblock \href {https://doi.org/10.1038/nclimate1068}
  {\path{doi:10.1038/nclimate1068}}.

\bibitem{LEE2021}
D.~Lee, D.~Fahey, A.~Skowron, M.~Allen, U.~Burkhardt, Q.~Chen, S.~Doherty,
  S.~Freeman, P.~Forster, J.~Fuglestvedt, A.~Gettelman, R.~{De León}, L.~Lim,
  M.~Lund, R.~Millar, B.~Owen, J.~Penner, G.~Pitari, M.~Prather, R.~Sausen,
  L.~Wilcox, The contribution of global aviation to anthropogenic climate
  forcing for 2000 to 2018, Atmospheric Environment 244 (2021) 117834.
\newblock \href {https://doi.org/10.1016/j.atmosenv.2020.117834}
  {\path{doi:10.1016/j.atmosenv.2020.117834}}.

\bibitem{UNTER2014}
S.~Unterstrasser, N.~Görsch, Aircraft-type dependency of contrail evolution,
  Journal of Geophysical Research: Atmospheres 119~(24) (2014) 14,015--14,027.
\newblock \href {https://doi.org/10.1002/2014JD022642}
  {\path{doi:10.1002/2014JD022642}}.

\bibitem{KAR2018}
B.~Kärcher, Formation and radiative forcing of contrail cirrus, Nat Commun 9
  (2018).
\newblock \href {https://doi.org/10.1038/s41467-018-04068-0}
  {\path{doi:10.1038/s41467-018-04068-0}}.

\bibitem{KAR1998}
B.~Kärcher, Physicochemistry of aircraft-generated liquid aerosols, soot, and
  ice particles: 1. {M}odel description, Journal of Geophysical Research:
  Atmospheres 103~(D14) (1998) 17111--17128.
\newblock \href {https://doi.org/10.1029/98JD01044}
  {\path{doi:10.1029/98JD01044}}.

\bibitem{WONG10}
H.-W. Wong, R.~C. Miake-Lye, Parametric studies of contrail ice particle
  formation in jet regime using microphysical parcel modeling, Atmospheric
  Chemistry and Physics 10~(7) (2010) 3261--3272.
\newblock \href {https://doi.org/10.5194/acp-10-3261-2010}
  {\path{doi:10.5194/acp-10-3261-2010}}.

\bibitem{BIER22}
A.~Bier, S.~Unterstrasser, X.~Vancassel, Box model trajectory studies of
  contrail formation using a particle-based cloud microphysics scheme,
  Atmospheric Chemistry and Physics 22~(2) (2022) 823--845.
\newblock \href {https://doi.org/10.5194/acp-22-823-2022}
  {\path{doi:10.5194/acp-22-823-2022}}.

\bibitem{SCHU98}
U.~Schumann, H.~Schlager, F.~Arnold, R.~Baumann, P.~Haschberger, O.~Klemm,
  Dilution of aircraft exhaust plumes at cruise altitudes, Atmospheric
  Environment 32~(18) (1998) 3097--3103.
\newblock \href {https://doi.org/10.1016/S1352-2310(97)00455-X}
  {\path{doi:10.1016/S1352-2310(97)00455-X}}.

\bibitem{PAOLI2003}
R.~Paoli, F.~Laporte, B.~Cuenot, T.~Poinsot, {Dynamics and mixing in jet/vortex
  interactions}, Physics of Fluids 15~(7) (2003) 1843--1860.
\newblock \href {https://doi.org/10.1063/1.1575232}
  {\path{doi:10.1063/1.1575232}}.

\bibitem{PAOLI2005}
R.~Paoli, F.~Garnier, Interaction of exhaust jets and aircraft wake vortices:
  small-scale dynamics and potential microphysical-chemical transformations,
  Comptes Rendus Physique 6~(4) (2005) 525--547, aircraft trailing vortices.
\newblock \href {https://doi.org/10.1016/j.crhy.2005.05.003}
  {\path{doi:10.1016/j.crhy.2005.05.003}}.

\bibitem{PAO2013}
R.~Paoli, L.~Nybelen, J.~Picot, D.~Cariolle, {Effects of jet/vortex interaction
  on contrail formation in supersaturated conditions}, Physics of Fluids 25~(5)
  (2013) 053305.
\newblock \href {https://doi.org/10.1063/1.4807063}
  {\path{doi:10.1063/1.4807063}}.

\bibitem{Lewe20}
D.~C. Lewellen, A {L}arge-{E}ddy {S}imulation study of contrail ice number
  formation, Journal of the Atmospheric Sciences 77~(7) (2020) 2585 -- 2604.
\newblock \href {https://doi.org/10.1175/JAS-D-19-0322.1}
  {\path{doi:10.1175/JAS-D-19-0322.1}}.

\bibitem{GUIGNERY12}
F.~Guignery, E.~Montreuil, O.~Thual, X.~Vancassel, Contrail microphysics in the
  near wake of a realistic wing through {RANS} simulations, Aerospace Science
  and Technology 23~(1) (2012) 399--408, 35th ERF: Progress in Rotorcraft
  Research.
\newblock \href {https://doi.org/10.1016/j.ast.2011.09.011}
  {\path{doi:10.1016/j.ast.2011.09.011}}.

\bibitem{Khou15}
J.-C. Khou, W.~Ghedha\"{\i}fi, X.~Vancassel, F.~Garnier, Spatial simulation of
  contrail formation in near-field of commercial aircraft, Journal of Aircraft
  52~(6) (2015) 1927--1938.
\newblock \href {https://doi.org/10.2514/1.C033101}
  {\path{doi:10.2514/1.C033101}}.

\bibitem{KHOU2016}
J.~Khou, W.~Ghedha{\"i}fi, X.~Vancassel, E.~Montreuil, F.~Garnier, {CFD}
  simulation of contrail formation in the near field of a commercial aircraft:
  Effect of fuel sulfur content, Meteorologische Zeitschrift 26~(6) (2017)
  585--596.
\newblock \href {https://doi.org/10.1127/metz/2016/0761}
  {\path{doi:10.1127/metz/2016/0761}}.

\bibitem{Wong14}
H.-W. Wong, M.~Jun, J.~Peck, I.~A. Waitz, R.~C. Miake-Lye, Detailed
  microphysical modeling of the formation of organic and sulfuric acid coatings
  on aircraft emitted soot particles in the near field, Aerosol Science and
  Technology 48~(9) (2014) 981--995.
\newblock \href {https://doi.org/10.1080/02786826.2014.953243}
  {\path{doi:10.1080/02786826.2014.953243}}.

\bibitem{Cantin25}
S.~Cantin, M.~Chouak, F.~Garnier, Effects of fuel sulfur content and nv{PM}
  emissions on contrail formation: a {CFD}-microphysics study including the
  role of organic compounds, Journal of Aerosol Science 188 (2025) 106612.
\newblock \href
  {https://doi.org/https://doi.org/10.1016/j.jaerosci.2025.106612}
  {\path{doi:https://doi.org/10.1016/j.jaerosci.2025.106612}}.

\bibitem{Yu24}
F.~Yu, B.~Kärcher, B.~E. Anderson, Revisiting contrail ice formation: impact
  of primary soot particle sizes and contribution of volatile particles,
  Environmental Science \& Technology 58~(40) (2024) 17650--17660, pMID:
  39323293.
\newblock \href {https://doi.org/10.1021/acs.est.4c04340}
  {\path{doi:10.1021/acs.est.4c04340}}.

\bibitem{GERZ1997}
T.~Gerz, T.~Ehret, Wingtip vortices and exhaust jets during the jet regime of
  aircraft wakes, Aerospace Science and Technology 1~(7) (1997) 463--474.
\newblock \href {https://doi.org/10.1016/S1270-9638(97)90008-0}
  {\path{doi:10.1016/S1270-9638(97)90008-0}}.

\bibitem{GAR1996}
F.~Garnier, L.~Jacquin, On the dynamics of engine jets behind a transport
  aircraft, Rep. AGARD CP-584 (1996).

\bibitem{JAC2007}
L.~Jacquin, P.~Molton, P.~Loiret, E.~Coustols, An experiment on jet-wake vortex
  interaction, AIAA Paper No. AIAA 2007-4363 (2007).
\newblock \href {https://doi.org/10.2514/6.2007-4363}
  {\path{doi:10.2514/6.2007-4363}}.

\bibitem{MAR2008}
P.~Margaris, D.~Marles, I.~Gursul, Experiments on jet/vortex interaction, Exp
  Fluids 44 (2008) 261--278.
\newblock \href {https://doi.org/10.1007/s00348-007-0399-7}
  {\path{doi:10.1007/s00348-007-0399-7}}.

\bibitem{BOELLE2023}
T.~Bölle, V.~Brion, M.~Couliou, P.~Molton, {Experiment on jet–vortex
  interaction for variable mutual spacing}, Physics of Fluids 35~(1) (2023)
  015117.
\newblock \href {https://doi.org/10.1063/5.0127634}
  {\path{doi:10.1063/5.0127634}}.

\bibitem{LABBE2007}
O.~Labbé, E.~Maglaras, F.~Garnier, Large-{E}ddy {S}imulation of a turbulent
  jet and wake vortex interaction, Computers \& Fluids 36~(4) (2007) 772--785.
\newblock \href {https://doi.org/10.1016/j.compfluid.2006.06.001}
  {\path{doi:10.1016/j.compfluid.2006.06.001}}.

\bibitem{PS2023}
P.~Saulgeot, V.~Brion, N.~Bonne, E.~Dormy, L.~Jacquin, Effects of atmospheric
  stratification and jet position on the properties of early aircraft
  contrails, Phys. Rev. Fluids 8 (2023) 114702.
\newblock \href {https://doi.org/10.1103/PhysRevFluids.8.114702}
  {\path{doi:10.1103/PhysRevFluids.8.114702}}.

\bibitem{RR2024}
J.~Ramsay, I.~Tristanto, S.~Shahpar, A.~John, {Assessing the environmental
  impact of aircraft/engine integration with respect to contrails}, Journal of
  Engineering for Gas Turbines and Power 146~(11) (2024) 111026.
\newblock \href {https://doi.org/10.1115/1.4066150}
  {\path{doi:10.1115/1.4066150}}.

\bibitem{YOU24}
Y.~Bouhafid, N.~Bonne, L.~Jacquin, Combined {R}eynolds-averaged
  {N}avier-{S}tokes/{L}arge-{E}ddy {S}imulations for an aircraft wake until
  dissipation regime, Aerospace Science and Technology 154 (2024) 109512.
\newblock \href {https://doi.org/10.1016/j.ast.2024.109512}
  {\path{doi:10.1016/j.ast.2024.109512}}.

\bibitem{REFLOCH}
A.~Refloch, B.~Courbet, A.~Murrone, P.~Villedieu, C.~Laurent, P.~Gilbank,
  J.~Troyes, L.~Tessé, G.~Chaineray, J.~Dargaud, E.~Quémerais, F.~Vuillot,
  {CEDRE} {S}oftware, AerospaceLab Journal (03 2011).

\bibitem{MEN93}
F.~R. Menter, Two-equation eddy-viscosity turbulence models for engineering
  applications, AIAA Journal 32~(8) (1994) 1598--1605.
\newblock \href {https://doi.org/10.2514/3.12149} {\path{doi:10.2514/3.12149}}.

\bibitem{WIL06}
D.~Wilcox, Turbulence Modeling for {CFD} ({T}hird Edition) ({H}ardcover), DCW
  Industries, 2006.

\bibitem{CHURCH09}
M.~J. Churchfield, G.~A. Blaisdell, Numerical simulations of a wingtip vortex
  in the near field, Journal of Aircraft 46~(1) (2009) 230--243.
\newblock \href {https://doi.org/10.2514/1.38086} {\path{doi:10.2514/1.38086}}.

\bibitem{FEFLO}
A.~Loseille, R.~Lohner, Anisotropic adaptive simulations in aerodynamics, 48th
  AIAA Aerospace Sciences Meeting Including the New Horizons Forum and
  Aerospace Exposition (2010).
\newblock \href {https://doi.org/10.2514/6.2010-169}
  {\path{doi:10.2514/6.2010-169}}.

\bibitem{LOSE11}
A.~Loseille, F.~Alauzet, Continuous mesh framework part {I}: well-posed
  continuous interpolation error, SIAM Journal on Numerical Analysis 49~(1)
  (2011) 38--60.
\newblock \href {https://doi.org/10.1137/090754078}
  {\path{doi:10.1137/090754078}}.

\bibitem{LOSp211}
A.~Loseille, F.~Alauzet, Continuous mesh framework part {II}: Validations and
  applications, SIAM Journal on Numerical Analysis 49~(1) (2011) 61--86.
\newblock \href {https://doi.org/10.1137/10078654X}
  {\path{doi:10.1137/10078654X}}.

\bibitem{MON2018}
E.~Montreuil, W.~Ghedhaifi, V.~Chmielaski, F.~Vuillot, F.~Gand, A.~Loseille,
  Numerical simulation of contrail formation on the {C}ommon {R}esearch {M}odel
  wing/body/engine configuration, AIAA 2018-3189 Atmospheric and Space
  Environments Conference (2018).
\newblock \href {https://doi.org/10.2514/6.2018-3189}
  {\path{doi:10.2514/6.2018-3189}}.

\bibitem{ALAU2019}
F.~Alauzet, L.~Frazza, 3{D} {RANS} anisotropic mesh adaptation on the high-lift
  version of {NASA}'s {C}ommon {R}esearch {M}odel {(HL-CRM)}, AIAA Aviation
  2019 Forum (2019).
\newblock \href {https://doi.org/10.2514/6.2019-2947}
  {\path{doi:10.2514/6.2019-2947}}.

\bibitem{YOU2023}
Y.~Bouhafid, N.~Bonne, L.~Jacquin, Aerodynamic simulation of the near field of
  an aircraft using different mesh adaptation strategies, Aerospace Europe
  Conference 2023 – 10TH {EUCASS} – 9TH {CEAS} (2023).
\newblock \href {https://doi.org/10.13009/EUCASS2023-421}
  {\path{doi:10.13009/EUCASS2023-421}}.

\bibitem{WONG08}
H.-W. Wong, P.~E. Yelvington, M.~T. Timko, T.~B. Onasch, R.~C. Miake-Lye,
  J.~Zhang, I.~A. Waitz, Microphysical modeling of ground-level
  aircraft-emitted aerosol formation: roles of sulfur-containing species,
  Journal of Propulsion and Power 24~(3) (2008) 590--602.
\newblock \href {https://doi.org/10.2514/1.32293} {\path{doi:10.2514/1.32293}}.

\bibitem{DAVIES76}
C.~Davies, Physics of drop formation in the atmosphere: by {Y}u {S}. {S}edunov
  (translated from russian), pp. x + 234. {H}ard cover. £8.40. {I}srael
  program for scientific translations, {J}ohn {W}iley \& {S}ons, {N}ew {Y}ork
  and {T}oronto. 1974, Journal of Aerosol Science 7~(2) (1976) 192.
\newblock \href {https://doi.org/10.1016/0021-8502(76)90075-6}
  {\path{doi:10.1016/0021-8502(76)90075-6}}.

\bibitem{KAR96}
B.~Kärcher, T.~Peter, U.~M. Biermann, U.~Schumann, The initial composition of
  jet condensation trails, Journal of Atmospheric Sciences 53~(21) (1996) 3066
  -- 3083.
\newblock \href
  {https://doi.org/10.1175/1520-0469(1996)053<3066:TICOJC>2.0.CO;2}
  {\path{doi:10.1175/1520-0469(1996)053<3066:TICOJC>2.0.CO;2}}.

\bibitem{JEN98}
E.~J. Jensen, O.~B. Toon, S.~Kinne, G.~W. Sachse, B.~E. Anderson, K.~R. Chan,
  C.~H. Twohy, B.~Gandrud, A.~Heymsfield, R.~C. Miake-Lye, Environmental
  conditions required for contrail formation and persistence, Journal of
  Geophysical Research: Atmospheres 103~(D4) (1998) 3929--3936.
\newblock \href {https://doi.org/10.1029/97JD02808}
  {\path{doi:10.1029/97JD02808}}.

\bibitem{PET07}
M.~D. Petters, S.~M. Kreidenweis, A single parameter representation of
  hygroscopic growth and cloud condensation nucleus activity, Atmospheric
  Chemistry and Physics 7~(8) (2007) 1961--1971.
\newblock \href {https://doi.org/10.5194/acp-7-1961-2007}
  {\path{doi:10.5194/acp-7-1961-2007}}.

\bibitem{CRM}
J.~Vassberg, M.~Dehaan, M.~Rivers, R.~Wahls, Development of a {C}ommon
  {R}esearch {M}odel for applied {CFD} validation studies, 26th AIAA Applied
  Aerodynamics Conference (2008).
\newblock \href {https://doi.org/10.2514/6.2008-6919}
  {\path{doi:10.2514/6.2008-6919}}.

\bibitem{JAC2001}
L.~Jacquin, D.~Fabre, P.~Geffroy, E.~Coustols, The properties of a transport
  aircraft wake in the extended near field - {A}n experimental study, 39th
  Aerospace Sciences Meeting and Exhibit (2001).
\newblock \href {https://doi.org/10.2514/6.2001-1038}
  {\path{doi:10.2514/6.2001-1038}}.

\bibitem{GAR1997}
F.~Garnier, S.~Brunet, L.~Jacquin, Modelling exhaust plume mixing in the near
  field of an aircraft, Annales Geophysicae 15 (1997).
\newblock \href {https://doi.org/10.1007/s00585-997-1468-1}
  {\path{doi:10.1007/s00585-997-1468-1}}.

\bibitem{Kar09}
B.~Kärcher, F.~Yu, Role of aircraft soot emissions in contrail formation,
  Geophysical Research Letters 36~(1) (2009).
\newblock \href {https://doi.org/https://doi.org/10.1029/2008GL036649}
  {\path{doi:https://doi.org/10.1029/2008GL036649}}.

\bibitem{Markl24}
R.~S. M\"arkl, C.~Voigt, D.~Sauer, R.~K. Dischl, S.~Kaufmann, T.~Harla{\ss},
  V.~Hahn, A.~Roiger, C.~Wei{\ss}-Rehm, U.~Burkhardt, U.~Schumann, A.~Marsing,
  M.~Scheibe, A.~D\"ornbrack, C.~Renard, M.~Gauthier, P.~Swann, P.~Madden,
  D.~Luff, R.~Sallinen, T.~Schripp, P.~Le~Clercq, Powering aircraft with
  100\,{\%} sustainable aviation fuel reduces ice crystals in contrails,
  Atmospheric Chemistry and Physics 24~(6) (2024) 3813--3837.
\newblock \href {https://doi.org/10.5194/acp-24-3813-2024}
  {\path{doi:10.5194/acp-24-3813-2024}}.

\bibitem{RIV11}
M.~Rivers, A.~Dittberner, Experimental investigations of the {NASA} {C}ommon
  {R}esearch {M}odel in the {NASA} {L}angley {N}ational {T}ransonic {F}acility
  and {NASA} {A}mes 11-{F}t {T}ransonic {W}ind {T}unnel ({I}nvited), 49th AIAA
  Aerospace Sciences Meeting including the New Horizons Forum and Aerospace
  Exposition (2011).
\newblock \href {https://doi.org/10.2514/6.2011-1126}
  {\path{doi:10.2514/6.2011-1126}}.

\bibitem{BALAK11}
S.~Balakrishna, M.~Acheson, Analysis of {NASA} {C}ommon {R}esearch {M}odel
  dynamic data, 49th AIAA Aerospace Sciences Meeting including the New Horizons
  Forum and Aerospace Exposition (2011).
\newblock \href {https://doi.org/10.2514/6.2011-1127}
  {\path{doi:10.2514/6.2011-1127}}.

\bibitem{SUSS01}
R.~Sussmann, K.~M. Gierens, Differences in early contrail evolution of
  two-engine versus four-engine aircraft: lidar measurements and numerical
  simulations, Journal of Geophysical Research: Atmospheres 106~(D5) (2001)
  4899--4911.
\newblock \href {https://doi.org/https://doi.org/10.1029/2000JD900533}
  {\path{doi:https://doi.org/10.1029/2000JD900533}}.

\bibitem{KOLO18}
D.~Kolomenskiy, R.~Paoli, Numerical simulation of the wake of an airliner,
  Journal of Aircraft 55~(4) (2018) 1689--1699.
\newblock \href {https://doi.org/10.2514/1.C034349}
  {\path{doi:10.2514/1.C034349}}.

\bibitem{CANTIN22}
S.~Cantin, M.~Chouak, F.~Morency, F.~Garnier, Eulerian–{L}agrangian
  {CFD}-microphysics modeling of a near-field contrail from a realistic
  turbofan, International Journal of Engine Research 23~(4) (2022) 661--677.
\newblock \href {https://doi.org/10.1177/1468087421993961}
  {\path{doi:10.1177/1468087421993961}}.

\bibitem{KLEI18}
J.~Kleine, C.~Voigt, D.~Sauer, H.~Schlager, M.~Scheibe, T.~Jurkat-Witschas,
  S.~Kaufmann, B.~Kärcher, B.~E. Anderson, In situ observations of ice
  particle losses in a young persistent contrail, Geophysical Research Letters
  45~(24) (2018) 13,553--13,561.
\newblock \href {https://doi.org/10.1029/2018GL079390}
  {\path{doi:10.1029/2018GL079390}}.

\bibitem{qext}
H.~C. Van~de Hulst, Light scattering by small particles. {N}ew {Y}ork ({J}ohn
  {W}iley and {S}ons), {L}ondon ({C}hapman and {H}all), 1957. {P}p. xiii, 470;
  103 {F}igs.; 46 {T}ables. 96s, Quarterly Journal of the Royal Meteorological
  Society 84~(360) (1958) 198--199.
\newblock \href {https://doi.org/10.1002/qj.49708436025}
  {\path{doi:10.1002/qj.49708436025}}.

\bibitem{Schu17}
U.~Schumann, R.~Baumann, D.~Baumgardner, S.~T. Bedka, D.~P. Duda,
  V.~Freudenthaler, J.-F. Gayet, A.~J. Heymsfield, P.~Minnis, M.~Quante,
  E.~Raschke, H.~Schlager, M.~V\'azquez-Navarro, C.~Voigt, Z.~Wang, Properties
  of individual contrails: a compilation of observations and some comparisons,
  Atmospheric Chemistry and Physics 17~(1) (2017) 403--438.
\newblock \href {https://doi.org/10.5194/acp-17-403-2017}
  {\path{doi:10.5194/acp-17-403-2017}}.

\end{thebibliography}

\end{document}